\documentclass[a4paper,11pt]{article}
\pdfoutput=1
\usepackage{jheppub}
\usepackage{graphicx}
\usepackage{dcolumn}
\usepackage{bm}
\usepackage{amssymb}
\usepackage{amsmath}
\usepackage{epsfig}    
\usepackage{color}
\usepackage{slashed}
\usepackage{hhline}
\usepackage{float}
\usepackage{csquotes} 
\usepackage{comment}
\usepackage{mathrsfs}
\usepackage{float}
\usepackage{appendix}

\def\be{\begin{equation}}
\def\ee{\end{equation}}
\newcommand{\bea}{\begin{eqnarray}}
\newcommand{\eea}{\end{eqnarray}}
\newcommand{\nn}{\nonumber}

\numberwithin{equation}{section}

\preprint{
\begin{minipage}{5cm}
\small
\flushright
KEK-TH-2567\\KYUSHU-HET-271
\end{minipage}}

\title{Novel moduli space in modular flavor models : a case study for modular $T'$ seesaw model with hidden $SU(2)$ gauge symmetry}

\author{Keiya Ishiguro$^{1}$,} 
\author{Takaaki Nomura$^{2}$,}
\author{Hiroshi Okada$^{3}$,}
\author{Yuta Orikasa$^{4}$,}
\author{Hajime Otsuka$^{3}$} 
\affiliation{
$^1$Graduate University for Advanced Studies (Sokendai), 1-1 Oho, Tsukuba, Ibaraki 305-0801, Japan\\
$^2$College of Physics, Sichuan University, Chengdu 610065, China \\
$^3$Department of Physics, Kyushu University, 744 Motooka, Nishi-ku, Fukuoka 819-0395, Japan\\
$^4$Institute of Experimental and Applied Physics, 
Czech Technical University in Prague, 
Husova 240/5, 110 00 Prague 1, Czech Republic\\
}
\emailAdd{ishigu@post.kek.jp}
\emailAdd{nomura@scu.edu.cn}
\emailAdd{okada.hiroshi@phys.kyushu-u.ac.jp}
\emailAdd{Yuta.Orikasa@utef.cvut.cz}
\emailAdd{otsuka.hajime@phys.kyushu-u.ac.jp}

\abstract{
We study flavor phenomenologies in a basis of a double covering of modular $A_4$ group
with a hidden $SU(2)$ symmetry, in which
we work on regions at nearby two fixed points and three special points.
These special points of $SL(2,\mathbb{Z})$ moduli space are statistically favored in flux compactifications of Type IIB string theory.
Neutrino masses are approximately obtained by using the seesaw mechanism due to our two additional symmetries.
We perform chi square numerical analysis for each of the fixed/special points in the normal and inverted hierarchies and demonstrate predictions for each case.
Then, we briefly discuss the possible signature of hidden gauge bosons at collider experiments from the hidden $SU(2)$ symmetry.

}
\makeatletter
\gdef\@fpheader{}
\makeatother

\begin{document}

\maketitle

\section{Introduction}

One of the main issues in the Standard Model (SM) is to explain the flavor structure of 
quarks and leptons by some mechanisms. 
Among many possibilities, a flavor symmetry plays an important role in understanding the hierarchical 
masses and mixing angles of the SM fermions. 


In models with flavor symmetries, Yukawa couplings are considered as a singlet or a non-trivial representation 
under a flavor symmetry. 
The former scenario was known to be the traditional approach to understand the flavor structure of 
quarks and leptons\footnote{See for more details, e.g., Refs. \cite{Altarelli:2010gt, Ishimori:2010au, Kobayashi:2022moq}.}, but the latter scenario is more attractive and natural from the viewpoint of ultraviolet physics \cite{Almumin:2021fbk}. 
Indeed, in the higher-dimensional theory such as the string theory, 
the Yukawa couplings are given by the overlap integral of matter wavefunctions 
once quarks and leptons live in the bulk space of the extra-dimensional space. 
It indicates that the Yukawa couplings will be controlled by a geometric symmetry of the extra-dimensional 
space such as the (finite part of) $SL(2,\mathbb{Z})$ modular symmetry in the single modulus 
and $Sp(2g,\mathbb{Z})$ modular symmetry in multi moduli case. 
The explicit model buildings have been performed in heterotic string on toroidal orbifolds \cite{Ferrara:1989qb,Lerche:1989cs,Lauer:1990tm,Baur:2019kwi,Baur:2019iai,Ishiguro:2021ccl} and Calabi-Yau threefolds \cite{Ishiguro:2020nuf,Ishiguro:2021ccl}, and Type IIB magnetized D-branes \cite{Kobayashi:2018rad,Kobayashi:2018bff,Ohki:2020bpo,Kikuchi:2020frp,Kikuchi:2020nxn,Almumin:2021fbk}. 


Since the Yukawa couplings depend on the so-called moduli fields, the moduli vacuum expectation values (VEVs) 
are important to predict the flavor pattern of quarks and leptons. 
In particular, residual symmetries in the $SL(2,\mathbb{Z})$ moduli space of $\tau$ exist at some particular points of moduli space, called fixed points, such as $\tau=i,\omega, i\infty$ with $\omega = \frac{-1+i\sqrt{3}}{2}$, corresponding to 
$\mathbb{Z}_2$, $\mathbb{Z}_3$ and $\mathbb{Z}_2$ symmetries, which was phenomenologically applied in the lepton sector \cite{Novichkov:2018ovf,Novichkov:2018yse,Novichkov:2018nkm,Ding:2019gof,Okada:2019uoy,King:2019vhv,Okada:2020rjb,Okada:2020ukr,Okada:2020brs,Feruglio:2021dte,Kobayashi:2021pav,Kobayashi:2022jvy}. 
Remarkably, the $SL(2,\mathbb{Z})$ modular symmetry breaking was performed in \cite{Kobayashi:2020hoc}, and moduli VEVs are predicted by a top-down approach, e.g., Type IIB string compactifications with background fluxes \cite{Ishiguro:2020tmo}, and the deviation of the moduli field from the fixed points was achieved in Ref. \cite{Ishiguro:2022pde}. 
In the top-down approach of Ref. \cite{Ishiguro:2020tmo}, distributions of moduli fields from a finite number of string flux vacua suggest us that the moduli VEVs are statistically favored at the other special points such as $\tau =  \frac{-1+i\sqrt{15}}{2}, \sqrt{3}i, \frac{-1+i\sqrt{15}}{4}$ in addition to the fixed points $\tau = i, \omega$. 
Note that these points are not protected by a certain symmetry. 


In the bottom-up approach, there is no guiding principle to pick up certain moduli VEVs from the $SL(2,\mathbb{Z})$ moduli space. 
In this paper, we discuss the phenomenological aspects of these special moduli values 
$\tau =  i, \omega,  \frac{-1+i\sqrt{15}}{2}, \sqrt{3}i, \frac{-1+i\sqrt{15}}{4}$, predicted by the top-down approach. 
Note that $\tau = i\infty$ will be excluded by the cancellation of D-brane charges, or in other words, we require the infinite value of background flux quanta to approach $\tau= i \infty$. 
For illustrative purposes, we focus on the lepton sector in the seesaw model with a double covering of modular $A_4$ symmetry; $T'$, and hidden $SU(2)$ gauge symmetry.
The hidden symmetry 
is useful to construct the seesaw model by discriminating heavy neutral fermions that typically appear in seesaw models in the framework of modular symmetry with an ordinal modular weight assignments~\cite{Kobayashi:2023qzt}.
The additional symmetries naturally appear in the low-energy effective action of the string theory~\footnote{See, Refs. \cite{Ferrara:1989qb,Lerche:1989cs}, for the realization of the modular $T'$ symmetry.}.
Even though the minimum flavor symmetry including an irreducible triplet is $A_4$ symmetry and a vast amount of literature has been appeared in refs.~\cite{Kobayashi:2018scp, Okada:2018yrn, Nomura:2019jxj, Okada:2019uoy, deAnda:2018ecu, Novichkov:2018yse, Nomura:2019yft, Okada:2019mjf,Ding:2019zxk, Nomura:2019lnr,Kobayashi:2019xvz,Asaka:2019vev,Zhang:2019ngf, Gui-JunDing:2019wap,Kobayashi:2019gtp,Nomura:2019xsb, Wang:2019xbo,Okada:2020dmb,Okada:2020rjb, Behera:2020lpd, Behera:2020sfe, Nomura:2020opk, Nomura:2020cog, Asaka:2020tmo, Okada:2020ukr, Nagao:2020snm, Okada:2020brs,Kang:2022psa, Nomura:2023usj,Mishra:2023ekx,Kumar:2023moh,CentellesChulia:2023zhu,Kashav:2022kpk}, in this paper, we apply its double covering group because of following reasons.
The first one is that this group possesses an irreducible doublet that could be identified as a representation for heavier Majorana fermions.
Therefore, the active neutrino mass matrix is minimally realized by rank two matrix.
The second one is that $T'$ allows us to 
use an odd number of modular weights for Yukawa couplings and different types of mass matrices could be obtained compared to the one with $A_4$.


In this paper, instead of hidden $U(1)$,
we adopt the non-Abelian gauge symmetry $SU(2)$ which also appears in the low-energy effective action of the string theory. \footnote{See, e.g., Ref. \cite{Gmeiner:2005vz}, for the statistical approach of D-brane model building on toroidal orientifolds and Ref. \cite{Otsuka:2018rki} for $SO(32)$ heterotic model building in general Calabi-Yau compactifications.} 
In a typical interpretation of "hidden symmetry", we expect {that} the {$SU(2)$} symmetry has to (almost) be separated from the SM sector {due to the structure of $SU(2)$}. It suggests that there exist no interactions between these two sectors or the interactions has to be very small.
However, if we introduce a hidden $U(1)$ symmetry, we have no way to eliminate the kinetic mixing between the SM and hidden sector except for assuming the mixing to be enough small by hand. In case of hidden non-Abelian symmetry, the mixing would not be appeared without an intermediate sector that has both the SM and hidden charge. The mixing can appear with the intermediate sector at a tree level or at a loop level~\cite{Nomura:2021tmi,Nomura:2021aep}. 
Moreover, a remnant symmetry after spontaneous symmetry breaking of $SU(2)$ controls the stability of dark matter even though we do not discuss here. The unbroken symmetry can be naturally preserved compared to the Abelian symmetry in which the $U(1)$ charge has to satisfy some artificial tuning~\cite{Krauss:1988zc}.  In fact, various applications of the hidden $SU(2)$ gauge symmetries have been studied in articles, e.g., a remaining $\mathbb{Z}_2$ symmetry in ref.~\cite{Ko:2020qlt}, $\mathbb{Z}_3(\mathbb{Z}_4)$ symmetry with a quadruplet(quintet) in Ref.~\cite{Chiang:2013kqa, Chen:2015nea, Chen:2015dea}, $\mathbb{Z}_2\times \mathbb{Z}'_2$ symmetry~\cite{Gross:2015cwa}, a custodial symmetry in refs.~\cite{Boehm:2014bia, Hambye:2008bq}, an unbroken $U(1)$ from $SU(2)$ in refs.~\cite{Baek:2013dwa,Khoze:2014woa,Daido:2019tbm}, a model adding hidden $U(1)_H$~\cite{Davoudiasl:2013jma} and a model with classical scale invariance~\cite{Karam:2015jta}. 
It is then challenging and interesting to consider neutrino mass generation with hidden $SU(2)$. 
Hidden $SU(2)$ can also be applied to discriminate heavy neutral fermions from the SM sector, and intermediate sector inducing mixing between hidden and the SM sector also play a role in generating neutrino mass.
In addition one finds that we need at least two generations of heavy neutral fermion if it is $SU(2)$ doublet in order to cancel the anomaly~\cite{Witten:1982fp}. This minimal two generations of hidden heavy doublet can be organized into $T'$ doublet and thus we have good compatibility between hidden $SU(2)$ and $T'$. 
In the model of this work, hidden $SU(2)_H$ symmetry is spontaneously broken by VEVs of scalar fields including $SU(2)_L \times SU(2)_H$ bidoublets. 
Then Majorana fermions in the hidden sector can mix with active neutrinos in the SM sector inducing the type-I like seesaw mechanism~\cite{Yanagida:1979gs, Minkowski:1977sc, Mohapatra:1979ia}.
Furthermore we would have interesting phenomenology from hidden vector bosons in the model.
In our scenario two hidden vector bosons from the $SU(2)_H$ mix with the SM neutral $Z$ boson since $SU(2)_L \times SU(2)_H$ bidoublet scalar fields get VEVs. On the other hand one hidden vector boson does not directly mix with the SM gauge boson. Thus we have two dark photons like vector bosons and one vector boson which does not couple to the SM particles. The latter one would be long lived one and could provide a distinguishable signature at the collider experiments. We discuss these hidden vector bosons and the possibility of their collider phenomenology briefly.

The remaining section is organized as follows. 
In Sect.2, we review two fixed points and three special points which are known as stable vacua and discuss their origin based on four-dimensional effective action of Type IIB string theory. In Sect.3, we show our model set up to generate the seesaw model in a framework of modular $T'$ and hidden gauge $SU(2)$ symmetries. Then, we demonstrate numerical analyses and display our predictions in each of the five points that we have discussed in Sect. 2. Then, we discuss the possibility of testability via colliders in the multiple dark gauge boson sector. Finally, we give our summary and discussions.

\section{Novel moduli values from top-down approach}
\label{sec:2}

In this section, we briefly review the distribution of moduli fields on the basis of four-dimensional effective action of Type IIB string theory. (For more details, see, e.g., Refs. \cite{Blumenhagen:2006ci,Ibanez:2012zz,Blumenhagen:2013fgp}.) 
Since the modulus field $\tau$ determines the fermion mass hierarchies and mixing angles in modular flavor models, it is important to fix the vacuum expectation value by some mechanisms. 
In higher-dimensional theories, background $p$-form fluxes in compact extra-dimensional spaces will induce the potential of moduli fields. Indeed, in Type IIB flux compactifications, the background three-form fluxes generate the four-dimensional superpotential \cite{Gukov:1999ya}:
\begin{align}
    W (S, \tau) = \int G_3 \wedge \Omega,
\end{align}
where $G_3 = F_3- SH_3$ is defined as a linear combination of Ramond-Ramond three-form $F_3$ and Neveu-Schwarz three-form $H_3$, and $\Omega$ is a holomorphic three-form of the internal six-dimensional space as a function of a certain complex structure modulus $\tau$. 
It results in the potential of axio-dilaton $S$ and complex structure modulus $\tau$. Specifically, we focus on $T^6/(\mathbb{Z}_2\times \mathbb{Z}_2)$ geometry which incorporating 
the overall $SL(2,\mathbb{Z})_\tau$ modular symmetry used in modular flavor models. 

On $T^6/(\mathbb{Z}_2\times \mathbb{Z}_2)$ geometry, the superpotential with the modular weight 3 is of the form:
\begin{align}
    W &= a^0 \tau^3- 3a\tau^2- 3b \tau-b_0 -S\left(c^0\tau^3-3c\tau^2-3d\tau-d_0 \right),
\label{eq:Wsim}
\end{align}
where the coefficients of the modulus $\tau$ represent a flux quanta. 
Together with the kinetic terms of $S$ and $\tau$ written in the K\"ahler potential:
\begin{align}
    K = -\ln (-i(S -\Bar{S})) -3 \ln\left(i (\tau-\bar{\tau})\right) -2\ln {\cal V},     
    \label{eq:Keff}
\end{align}
one can check the $PSL(2,\mathbb{Z})_\tau$ and $PSL(2,\mathbb{Z})_S$ modular symmetries 
following Refs. \cite{Betzler:2019kon,Ishiguro:2020tmo}. 
Throughout this paper, we do not study the stabilization of volume moduli ${\cal V}$, 
but it is possible to stabilize them following the prescription of Ref. \cite{Ishiguro:2022pde} in the context of modular flavor models. 
The moduli $S$ and $\tau$ can be stabilized at supersymmetric minima:
\begin{align}
    \partial_S W = \partial_\tau W = W = 0,
\end{align}
leading to the following vacuum expectation value of moduli fields:
\begin{align}
    S &=  \frac{ r \tau + s}{u \tau + v},
\nonumber\\
    \tau &=
    \left\{
    \begin{array}{c}
\frac{ - m + \sqrt{m^2 - 4 l n}}{2 l} \quad (l, n > 0)
    \\
    \frac{ - m - \sqrt{m^2 - 4 l n}}{2 l} \quad (l, n < 0)
\end{array}
\right.
.
\end{align}
Here, we redefine the flux quanta
\begin{alignat}{4}
    r l &=  a^0,~ &  r m + s l &=  -3 a,~ &  r n + s m  &= -3 b,~ &  s n &=  -b_0, \nonumber \\ 
    u l &= c^0,~ &  u m + v l &= -3 c,~  &  u n + v m &= -3 d,~ &  v n &= -d_0.
\end{alignat}

It was known that fixed points of $SL(2,\mathbb{Z})_\tau$ moduli space $\tau = \omega, i$ with $\omega =( -1+i\sqrt{3})/2$ are statistically favored in the flux landscape. Note that a region around the remaining fixed point $\tau = i\infty$ cannot be realized in flux compactifications due to the fact that the D3-brane charge induced by three-form flux quanta:
\begin{align}
    N_{\rm flux}&= \int H_3\wedge F_3 = 
    c^0b_0 -d_0a^0 +3(cb -da)
\end{align}
should be canceled in a compact space, and the value is upper bounded by D-brane and O-plane charges in string compactifications. 
Remarkably, the flux landscape also prefer other special points of $SL(2,\mathbb{Z})$ moduli space:
\begin{align}
    \tau = \frac{-1+i\sqrt{15}}{2}, \sqrt{3}i, \frac{-1+i\sqrt{15}}{4}, \cdots,
\end{align}
as shown in Table \ref{tab:Nflux10}. 
Here, we list the stable vacua, up to 5, in the descending order of the probability by setting a maximum value of $N_{\rm flux}$, that is, $N_{\rm flux}^{\rm max}=192\times 10$. 
Although the value of probabilities depends on $N_{\rm flux}^{\rm max}$, the statistical behavior is similar 
for other $N_{\rm flux}^{\rm max}$. 
In this respect, it is interesting to figure out a phenomenological aspect of these novel points in the moduli space. 
In the following sections, we deal with a modular $T'$ seesaw model with hidden $SU(2)$ gauge symmetry 
as a case study.

\begin{table}[H]
\centering
\begin{tabular}{|c|c|c|c|c|c|} \hline
$({\rm Re}\,\tau, {\rm Im}\,\tau)$ & ($-\frac{1}{2}, \frac{\sqrt{3}}{2}$) & ($0, \sqrt{3}$) & ($-\frac{1}{2}, \frac{\sqrt{15}}{2}$) 
 & ($-\frac{1}{4}, \frac{\sqrt{15}}{4}$) &($0, 1$)\\ \hline
Probability ($\%$) & 62.3 & 7.55 & 7.55 & 7.55 & 5.66\\ \hline
\end{tabular}
\caption{Probabilities of the stable vacua as functions of $({\rm Re}\,\tau, {\rm Im}\,\tau)$ 
in the finite number of flux vacua \cite{Ishiguro:2020tmo}.} 
\label{tab:Nflux10}
\end{table}

\section{Model setup}
\label{sec:3}

\begin{table}[h]
\centering
\begin{tabular}{|c||c|c|c||c|c|c|c|c|c|}\hline
& ~$\hat L$~& ~$(\hat e^c,\hat \mu^c,\hat \tau^c)$~& ~$\hat\Sigma^c$~& ~$\hat\Phi_1$~& ~$\hat\Phi_2$~ & ~$\hat H_X$~& ~$\hat H_u$~& ~$\hat H_d$~& ~$\hat\Delta$~\\\hline\hline 
$SU(2)_H$  & $\bm{1}$  &  $\bm{1}$ & $\bm{2}$ & $\bm{2}$& $\bm{2}$ & $\bm{2}$ & $\bm{1}$   &  $\bm{1}$ &
 $\bm{3}$  \\ \hline
$SU(2)_L$   & $\bm{2}$  & $\bm{1}$  & $\bm{1}$  & $\bm{2}$& $\bm{2}$  & $\bm{1}$ & $\bm{2}$ & $\bm{2}$ & $\bm{1}$   \\\hline 
$U(1)_Y$    & $-\frac12$  & $1$ & $0$  & $\frac12$& $-\frac12$  & $0$ & $\frac12$   &{$-\frac12$}   & $0$  \\\hline
$T'$   & $\bm{3}$  & $(\bm{1,1',1''})$  & $\bm{2}$  & $\bm{1}$& $\bm{1}$  & $\bm{2}$ & $\bm{1}$ & $\bm{1}$ & $\bm{1}$   \\\hline 
$-k_I$   & $-2$  & $(-4,-4,-8)$  & $-{1}$  & $-4$& $-4$  & $-1$ & $0$ & $0$ & $-4$   \\\hline 
\end{tabular}
\caption{Charge assignments of matter superfields 
under $SU(2)_H \times SU(2)_L \times U(1)_Y \times T'$, where all of them are singlet under $SU(3)_C$ and R-parity is assigned; $\hat L,\hat e,\hat\Sigma^c$ for minus and the other superfields for plus. 
$\hat \Sigma^c$ has two generations that are organized in $T'$ doublet in order to cancel the anomaly for $SU(2)_H$~\cite{Witten:1982fp}. }\label{tab:1}
\end{table}

In this section we formulate our model in which we introduce hidden $SU(2)_H$ gauge symmetry and modular $T'$ symmetry. The matter superfields in Lepton and scalar sector are summarized in Table~\ref{tab:1} with their charge assignment under $SU(2)_H \times SU(2)_L \times U(1)_Y \times T'$ and modular weight where scalar sector means the scalar components in the sector develops VEVs. 
In the scalar sector, we introduce $SU(2)_H$ doublet $H_X$ and $\Phi_{1,2}$ where the former one is SM gauge singlet and the latter one is also $SU(2)_L$ doublet with $U(1)_Y$ charge $1/2$.
In our scenario all these scalar fields develop VEVs inducing spontaneous symmetry breaking.
The scalar components in the superfields are written as follows:
\begin{align}
& H_u = \begin{pmatrix} h_u^+ \\ \frac{1}{\sqrt{2}} (v_u + h_u + i a_u) \end{pmatrix}, 
\ 
H_d = \begin{pmatrix} \frac{1}{\sqrt{2}} (v_d + h_d + i a_d)  \\ h_d^-\end{pmatrix}, \\
& H_X = \begin{pmatrix} \eta_{1/2} \\ \frac{1}{\sqrt{2}} (v_X + p_{-1/2} + i q_{-1/2})\end{pmatrix}, \\
& \Phi_1 = \begin{pmatrix} \phi_{1/2}^+ & \phi'^+_{-1/2} \\ 
\frac{1}{\sqrt{2}} (\kappa + r_{1/2} + i a_{1/2}) & \frac{1}{\sqrt{2}} (\kappa' + r'_{-1/2} + i a'_{-1/2}) \end{pmatrix},\\
& \Phi_2 = \begin{pmatrix}  \frac{1}{\sqrt{2}} (\zeta + q_{1/2} + i b_{1/2}) & \frac{1}{\sqrt{2}} (\zeta' + q'_{-1/2} + i b'_{-1/2})\\
 \varphi^-_{1/2} & \varphi'^-_{-1/2}  \end{pmatrix},\\
& \Delta = \begin{pmatrix} \frac{1}{\sqrt{2}}\sigma_0 & v_+ + \sigma_{+1}\\
 v_- + \sigma_{-1} & - \frac{1}{\sqrt{2}}\sigma_0  \end{pmatrix},
\end{align}
where $v_{u,d,X}$ and $\kappa,\kappa',\zeta,\zeta'$ are VEVs for corresponding fields, and the upper number indices are electric charges and the lower ones are hidden charges. Notice here that fields have zero charges without number symbols, and they have plus charges under R-parity.
In addition, $SU(2)_H$ doublet fermions $\Sigma^c$ are introduced which is taken as right-handed and SM gauge singlet.
We write the fermionic particles in the superfield $\Sigma^c$ with their components as
\begin{equation}
\Sigma^c_i = \begin{pmatrix} N^{ic}_{1/2} \\ n^{ic}_{-1/2} \end{pmatrix},
\end{equation}
where both component fields are electrically neutral, and the field has to have the even number of generations for guaranteeing the theory to be anomaly free associated with $SU(2)_H$; $i=1,2$~\cite{Witten:1982fp}. 
In the model, we fix two generations to this field as minimal choice and they are organized as $T'$ doublet. 
$\hat \Sigma^c$ has minus R-parity charge as well as SM leptons.
Then, the invariant superpotential under these symmetries is given by
\begin{align}
W &= a_\ell [Y^{(6)}_{3_1}  e^c  \hat L \cdot \hat H_d]
+
a'_\ell [Y^{(6)}_{3_2}  e^c \hat L \cdot \hat H_d] \nonumber \\
&
+
b_\ell [Y^{(6)}_{3_1}  \mu^c \hat L \cdot \hat H_d]
+
b'_\ell [Y^{(6)}_{3_2}  \mu^c \hat L \cdot \hat H_d] \nonumber \\
&+
c_\ell [Y^{(10)}_{3_1}  \tau^c \hat L \cdot \hat H_d]
+
c'_\ell [Y^{(10)}_{3_2}  \tau^c \hat L \cdot \hat H_d]
+
c''_\ell [Y^{(10)}_{3_3}  \tau^c \hat L \cdot \hat H_d] \nonumber \\
&+  a_\nu [Y^{(7)}_{2_1}  \hat L\cdot  \hat \Phi_1 \cdot  \hat\Sigma^c]
+
  a'_\nu [Y^{(7)}_{2_2}  \hat L\cdot  \hat \Phi_1 \cdot  \hat\Sigma^c]\nn\\
&+
b_\nu [Y^{(7)}_{2'}  \hat L\cdot  \hat \Phi_1 \cdot  \hat\Sigma^c]
+
c_\nu [Y^{(7)}_{2''}  \hat L\cdot  \hat \Phi_1 \cdot  \hat\Sigma^c] \nonumber \\
&+\mu_H \hat H_u\cdot \hat H_d
+ \kappa_{\Sigma_1} Y^{(6)}_{3_1} \hat \Sigma^c \cdot \hat \Delta  \hat \Sigma^c 
+ \kappa_{\Sigma_2} Y^{(6)}_{3_2} \hat \Sigma^c \cdot \hat \Delta  \hat \Sigma^c 
+ \mu_\Delta \hat \Delta \hat \Delta
+ \mu_\Phi \hat\Phi_1 \cdot \hat\Phi_2 \nn\\
&+ a \hat \Phi_1 \hat \Delta \cdot \hat \Phi_2+ b \hat H_d\cdot \hat \Phi_1\cdot \hat H_X 
+ c \hat H_u \cdot \hat \Phi_2 \cdot \hat H_X 
+ d \hat H_X \cdot \hat \Delta \cdot \hat H_X
, \label{eq:Lfermion}
\end{align}
 where $\cdot\equiv i\sigma_2$ is the second Pauli matrix, $[\cdots]$ represents the true singlet under modular $T'$ symmetry, and bare mass term $ \hat \Sigma^c \cdot \hat \Sigma^c$ vanishes. 
 Note here that $\mu_{\Sigma}$ should be anti-symmetric matrix due to anti-symmetric contraction of $SU(2)_H$ indices in the term.
It suggests that $\mu_{\Sigma}$ reduces the matrix rank by one, and we cannot formulate the active neutrino mass matrix. Thus, we introduce $\hat\Delta$ that leads to the term $\kappa_\Sigma$ as we will see later. The bi-doublet plays a role in inducing the Dirac mass terms between active neutrino and hidden Majorana fermions that are also needed to construct the neutrino mass matrix.
The explicit forms of modular forms are summarized in the appendix.
We then obtain mass matrices from terms in the superpotential by expanding them using formulas for products of $T'$ representations. 

In the model charged lepton mass matrix is obtained from first 7 terms in the superpotential after $H_d$ developing its VEV.
The mass matrix is explicitly written by
\begin{align}
M_\ell = & \frac{v_d}{\sqrt2} 
\begin{pmatrix} 
a_\ell& 0 & 0 \\
0 & b_\ell& 0\\
0 & 0 & c_\ell
\end{pmatrix} \nonumber \\ & \quad \times
\begin{pmatrix}
y_1^{(6)} + \epsilon_e y'^{(6)}_1 & y_3^{(6)} + \epsilon_e y^{(6)}_3 & y_2^{(6)} + \epsilon_e y'^{(6)}_2 \\
y_3^{(6)} + \epsilon_\mu y'^{(6)}_1 & y_2^{(6)} + \epsilon_\mu y'^{(6)}_2 & y_1^{(6)} + \epsilon_\mu y'^{(6)}_1 \\
y_2^{(6)} + \epsilon_\tau y'^{(10)}_2 + \epsilon'_\tau y''^{(10)}_2 & y_1^{(6)} + \epsilon_\tau y'^{(10)}_1 + \epsilon'_\tau y''^{(10)}_1 & y_3^{(6)} + \epsilon_\tau y'^{(10)}_3 + \epsilon'_\tau y''^{(10)}_3
\end{pmatrix},
\end{align}
where $\{\epsilon_e, \epsilon_\mu, \epsilon_\tau, \epsilon'_\tau \} = \{a'_\ell/a_\ell, b'_\ell/b_\ell, c'_\ell/c_\ell, c''_\ell/c_\ell\}$.
We diagonalize the matrix to obtain the charged-lepton mass eigenvalues as ${\rm  diag}( |m_e|^2, |m_\mu|^2, |m_\tau|^2)\equiv V_{e_L}^\dag m^\dag_\ell m_\ell V_{e_L} $ where $V_{eL}$ is a unitary matrix. In our numerical analysis we fix three input parameters $\{a_\ell, b_\ell, c_\ell \}$ by 
solving following relations:
\begin{align}
&{\rm Tr}[m_\ell {m_\ell}^\dag] = |m_e|^2 + |m_\mu|^2 + |m_\tau|^2,\\
&{\rm Det}[m_\ell {m_\ell}^\dag] = |m_e|^2  |m_\mu|^2  |m_\tau|^2,\\
&({\rm Tr}[m_\ell {m_\ell}^\dag])^2 -{\rm Tr}[(m_\ell {m_\ell}^\dag)^2] =2( |m_e|^2  |m_\mu|^2 + |m_\mu|^2  |m_\tau|^2+ |m_e|^2  |m_\tau|^2 ).
\end{align}
Here, we apply experimentally measured charged-lepton masses summarized in PDG~\cite{ParticleDataGroup:2018ovx}.

After spontaneous symmetry breaking, we obtain a mass matrix for neutral fermion including the SM neutrino such that
\begin{equation}
M =  \begin{pmatrix} {\bf 0}_{3 \times 3} & M_{\nu n} & M_{\nu N} \\ M^T_{\nu n} & M_{n} & {\bf 0}_{2 \times 2} \\ M^T_{\nu N} & {\bf 0}_{2 \times 2} & M_{N} \end{pmatrix} + ({\rm transposed}),
\end{equation}
where the basis is $[\nu, n^c_{-1/2}, N^c_{-1/2}]$.
The element $M_{\nu n (\nu N)}$ is $3\times2$ matrix that is obtained from terms with coefficient $\{a_\nu, a'_\nu, b_ \nu, c_\nu \}$ in superpotential after bidoublet $\Phi_1$ develops its VEV.
Explicit form of elements in $M_{\nu n (\nu N)}$ are written by
\begin{align}
M_{\nu n[\nu N]} & = \frac{\kappa [\kappa']}{\sqrt2} 
\begin{pmatrix} 
\sqrt{2} b_\nu e^{\frac{5 \pi}{12} i} f'^{(7)}_1 - c_\nu f''^{(7)}_2 & \sqrt{2} e^{\frac{7 \pi}{12} i}( a_\nu f^{(7)}_2  + a'_\nu g^{(7)}_2) - c_\nu f''^{(7)}_1 \\ \sqrt{2} e^{\frac{5 \pi}{12}}( a_\nu f^{(7)}_1  + a'_\nu g^{(7)}_1) - b_\nu f'^{(7)}_2  & - b_\nu f'^{(7)}_1 +\sqrt{2} e^{\frac{7 \pi}{12} i} c_\nu f''^{(7)}_2 \\  - a_\nu f_2^{(7)} - a'_\nu g_2^{(7)} + \sqrt{2} c_\nu e^{ \frac{5 \pi}{12} i} f''^{(7)}_1 & - a_\nu f_1^{(7)} - a'_\nu g_1^{(7)} + \sqrt{2} b_\nu e^{\frac{7 \pi}{12} i} f'^{(7)}_2  
\end{pmatrix}
\nonumber \\
& \equiv \frac{\kappa [\kappa']}{\sqrt2} \tilde{M}_{\nu n}.  
\end{align}
The $2\times 2$ matrices $M_n$ and $M_N$ are obtained from the terms with $\{ \kappa_{\Sigma_1}, \kappa_{\Sigma_2} \}$ coefficient in the superpotential after triplet $\Delta$ developing a VEV.
We can explicitly write them such that
\begin{align}
M_{N[n]} & = v_-[-v_+] 
\begin{pmatrix} 
\kappa_{\Sigma_1} y_2^{(6)} + \kappa_{\Sigma_2} y'^{(6)}_2 & \frac{1}{\sqrt{2}} e^{\frac{7 \pi}{12} i} (\kappa_{\Sigma_1} y_3^{(6)} + \kappa_{\Sigma_2} y'^{(6)}_3) \\
\frac{1}{\sqrt{2}} e^{\frac{7 \pi}{12} i} (\kappa_{\Sigma_1} y_3^{(6)} + \kappa_{\Sigma_2} y'^{(6)}_3) & e^{\frac{\pi}{6} i} (\kappa_{\Sigma_1} y_1^{(6)} + \kappa_{\Sigma_2} y'^{(6)}_1)
\end{pmatrix} \nonumber \\ 
& \equiv v_{-}[-v_+] \tilde{M}_{n}.
\end{align}
The active neutrino mass in the mode is given by type-I seesaw like with both Majorana fermion $n$ and $N$ such that 
\begin{align}
m_\nu & \simeq - M_{\nu n} M_n^{-1} M_{\nu n}^T - M_{\nu N} M_N^{-1} M_{\nu N}^T, \nonumber \\
& \left(\frac{\kappa^2}{2v_+} -\frac{\kappa'^2}{2v_-} \right) \tilde{M}_{\nu n} \tilde{M}_n^{-1} \tilde{M}_{\nu n}^T
\equiv \epsilon \tilde m_\nu,
\end{align}
where $\epsilon\equiv  \frac{\kappa^2}{2v_+} - \frac{\kappa'^2}{2v_-}$. 
$m_\nu$ is diagonalized by a unitary matrix $V_{\nu}$; $D_\nu=|\epsilon| \tilde D_\nu= V_{\nu}^T m_\nu V_{\nu}=|\epsilon| V_{\nu}^T \tilde m_\nu V_{\nu}$.
Then $|\epsilon|$ is determined by
\begin{align}
(\mathrm{NH}):\  |\epsilon|^2= \frac{|\Delta m_{\rm atm}^2|}{\tilde D_{\nu_3}^2-\tilde D_{\nu_1}^2},
\quad
(\mathrm{IH}):\  |\epsilon|^2= \frac{|\Delta m_{\rm atm}^2|}{\tilde D_{\nu_2}^2-\tilde D_{\nu_3}^2},
 \end{align}
where $\Delta m_{\rm atm}^2$ is the atmospheric neutrino mass square difference. (NH) and (IH) respectively represent the normal hierarchy and the inverted hierarchy. 
Then, the solar mass square difference is written in terms of $|\epsilon|$ as follows:
\begin{align}
\Delta m_{\rm sol}^2=  |\epsilon|^2 ({\tilde D_{\nu_2}^2-\tilde D_{\nu_1}^2}),
\end{align}
 which is compared to the observed value in our numerical analysis.
 %
The observed mixing matrix is defined by $U=V_{e_L}^\dag V_\nu$~\cite{Maki:1962mu}, where
it is parametrized by the following three mixing angles, one CP violating Dirac phase $\delta_{\rm CP}$,
and one Majorana phase $\alpha_{21}$:
\begin{equation}
U = 
\begin{pmatrix} c_{12} c_{13} & s_{12} c_{13} & s_{13} e^{-i \delta_{\rm CP}} \\ 
-s_{12} c_{23} - c_{12} s_{23} s_{13} e^{i \delta_{\rm CP}} & c_{12} c_{23} - s_{12} s_{23} s_{13} e^{i \delta_{\rm CP}} & s_{23} c_{13} \\
s_{12} s_{23} - c_{12} c_{23} s_{13} e^{i \delta_{\rm CP}} & -c_{12} s_{23} - s_{12} c_{23} s_{13} e^{i \delta_{\rm CP}} & c_{23} c_{13} 
\end{pmatrix}
\begin{pmatrix} 1 & 0 & 0 \\ 0 & e^{i \frac{\alpha_{21}}{2}} & 0 \\ 0 & 0 & 1 \end{pmatrix},
\end{equation}
where $c_{ij}$ and $s_{ij}(i,j=1,2,3; i < j)$ stand for $\cos \theta_{ij}$ and $\sin \theta_{ij}$, respectively. 
Each of the mixings is given in terms of the component of $U$ as follows:
\begin{align}
\sin^2\theta_{13}=|U_{e3}|^2,\quad 
\sin^2\theta_{23}=\frac{|U_{\mu3}|^2}{1-|U_{e3}|^2},\quad 
\sin^2\theta_{12}=\frac{|U_{e2}|^2}{1-|U_{e3}|^2},
\end{align}
and the Majorana phase $\alpha_{21}$ and Dirac phase $\delta_{\rm CP}$ are found in terms of the following relations:
\begin{align}
&
 \text{Im}[U^*_{e1} U_{e2}] = c_{12} s_{12} c_{13}^2 \sin \left( \frac{\alpha_{21}}{2} \right), \
 \text{Im}[U^*_{e1} U_{e3}] = - c_{12} s_{13} c_{13} \sin  \delta_{\rm CP} 
,\\
&
 \text{Re}[U^*_{e1} U_{e2}] = c_{12} s_{12} c_{13}^2 \cos \left( \frac{\alpha_{21}}{2} \right), \
 \text{Re}[U^*_{e1} U_{e3}] = c_{12} s_{13} c_{13} \cos  \delta_{\rm CP} 
,
\end{align}
where $\alpha_{21}/2,\ \delta_{\rm CP}$
are subtracted from $\pi$, when $\cos\left(\frac{\alpha_{21}}2\right),\ \cos\delta_{\rm CP}$ are negative.
In addition, the effective mass for the neutrinoless double beta decay is given by
\begin{align}
\langle m_{ee}\rangle=|\epsilon||\tilde D_{\nu_1} \cos^2\theta_{12} \cos^2\theta_{13}+\tilde D_{\nu_2} \sin^2\theta_{12} \cos^2\theta_{13}e^{i\alpha_{21}}+\tilde D_{\nu_3} \sin^2\theta_{13}e^{-2i\delta_{\rm CP}}|,
\end{align}
where we can compare our prediction with experimental measurements by current/future neutrinoless beta decay experiments such as KamLAND-Zen~\cite{KamLAND-Zen:2016pfg, KamLAND-Zen:2022tow}, nEXO~\cite{nEXO:2017nam} and LEGEND~\cite{LEGEND:2017cdu}.

\section{Numerical analysis}
In this section, we demonstrate the allowed space, applying $\chi$ square numerical analysis to satisfy the current neutrino oscillation data at around the each of the fixed/special points of $\tau$, where we randomly select within the ranges of input dimensionless parameters as
\begin{equation}
\{\epsilon_e, \epsilon_\mu, \epsilon_\tau, \epsilon'_\tau, a_\nu, a'_\nu, b_\nu, c_\nu, \kappa_{\Sigma_{1,2}} \} \in [10^{-6}-10^3]
\end{equation}
where parameters $\{a_\ell, b_\ell, c_\ell \}$ are determined to obtain charged lepton masses. 
Then, we take into consideration of five reliable measured neutrino oscillation data; ($\Delta m^2_{\rm atm}$, $\Delta m^2_{\rm sol}$, $\sin^2\theta_{13}$, $\sin^2\theta_{23}$, $\sin^2\theta_{12}$)~\cite{Esteban:2018azc}, while $\delta_{\rm CP}$ be an output parameter due to large ambiguity of experimental result in $3\sigma$ interval. 
For the modulus $\tau$, we first fix it at the fixed and special points and search for the free parameters satisfying observed data. 
If we could not find the allowed region, we next consider $\tau$ around the fixed/special points.

\subsection{NH}

Here we summarize our results for the NH case at each of the fixed/special points.

\subsubsection{$\tau=i$}

\begin{figure}[H]
  \includegraphics[scale=0.4]{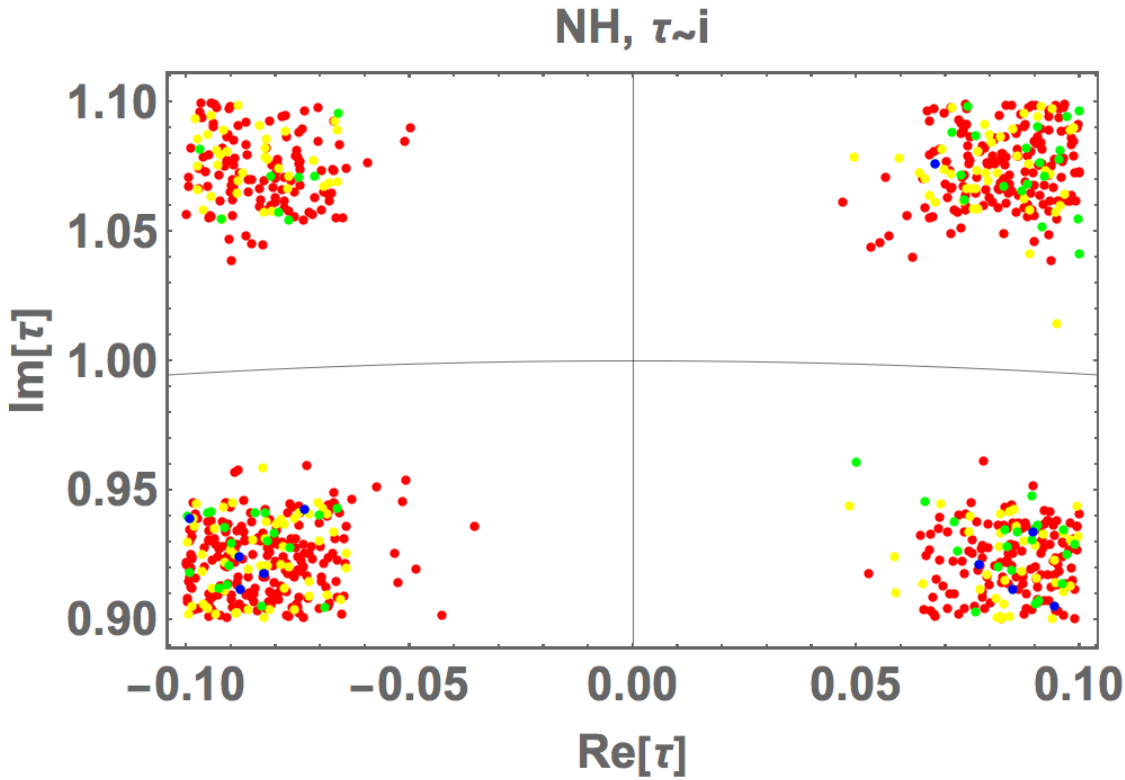}
  \includegraphics[scale=0.4]{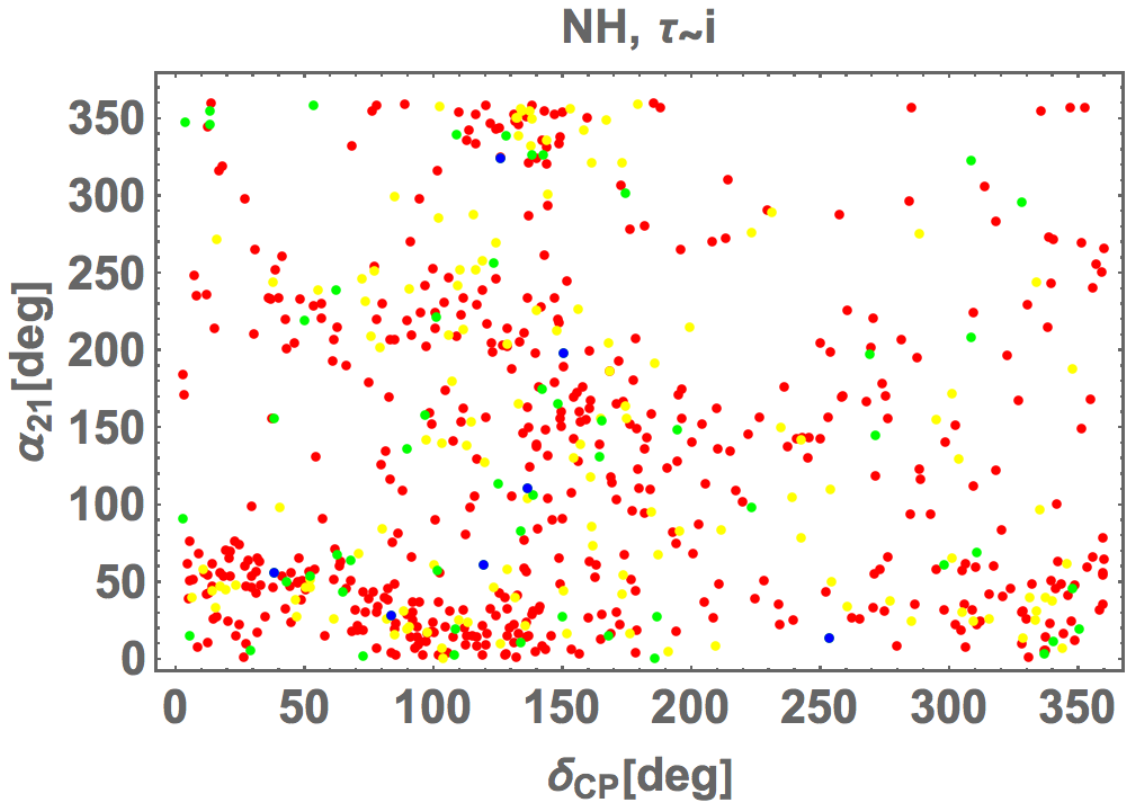}\\
   \includegraphics[scale=0.4]{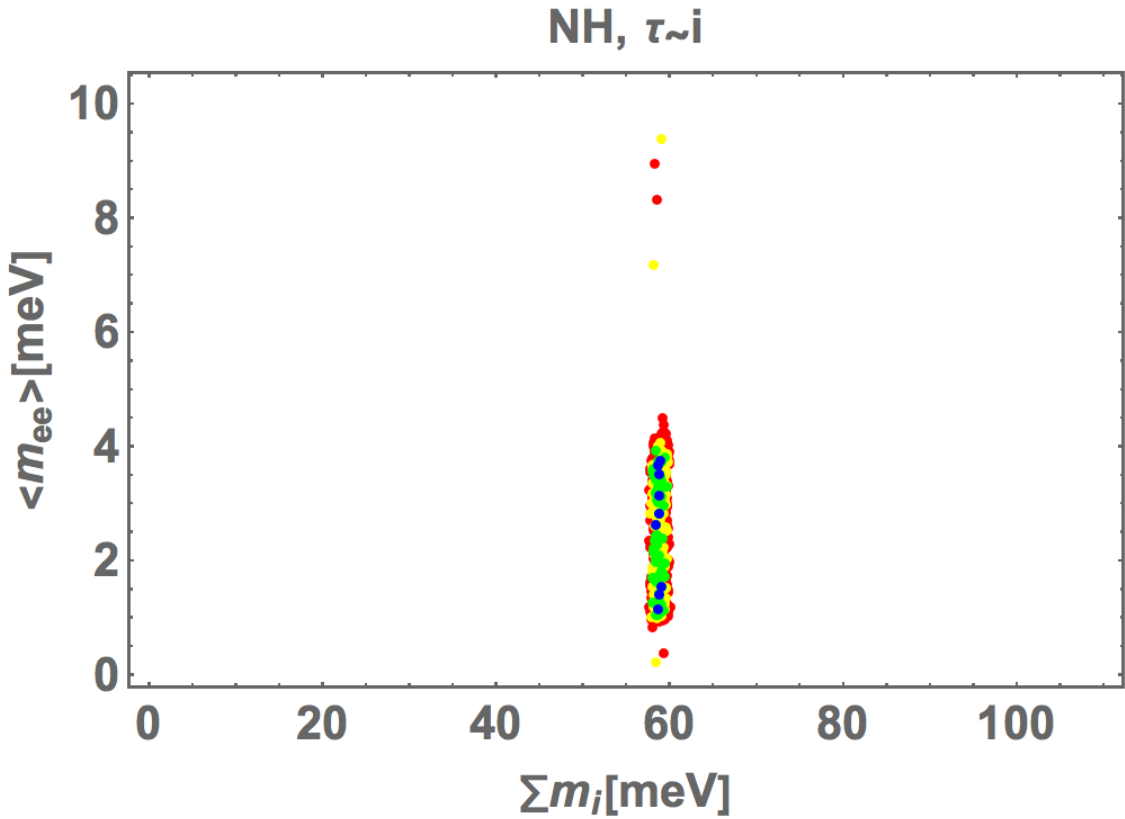}
   \includegraphics[scale=0.4]{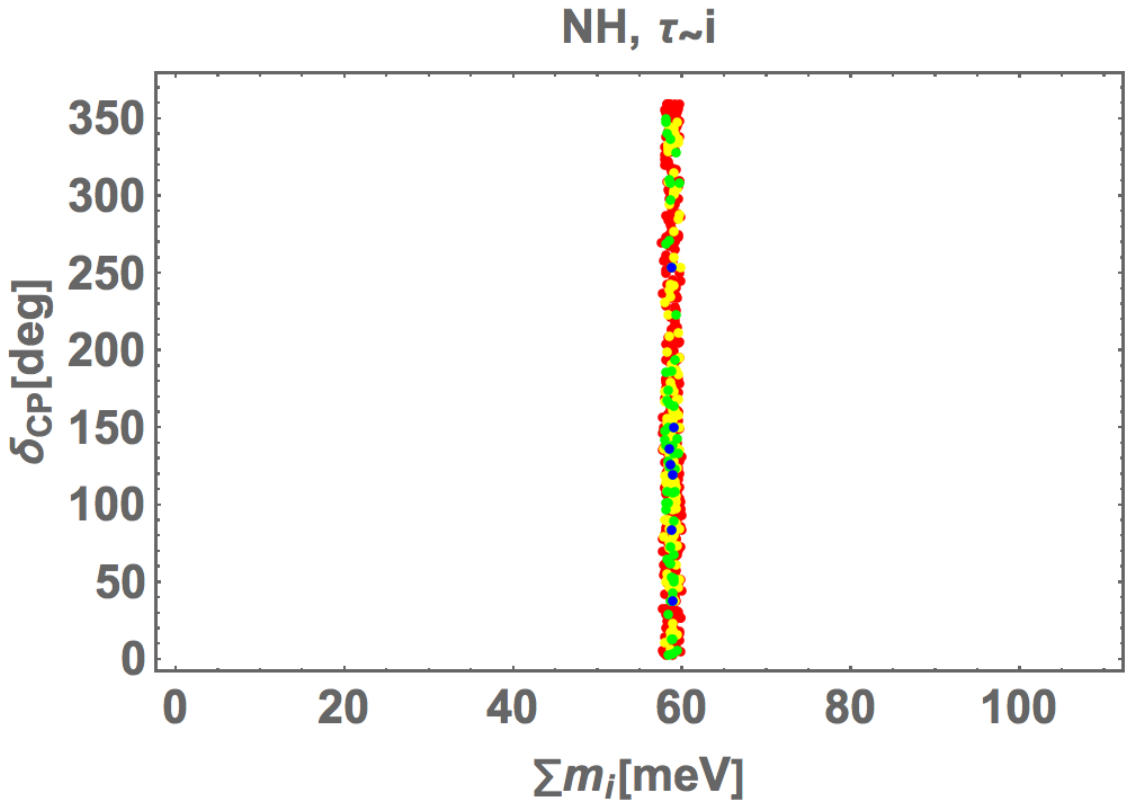}
 \caption{NH: Allowed regions in case of $\tau=i$. The colors of points indicate the $\chi^2$ result corresponding to the level of fitting as follows; blue: within $1\sigma$ level, green: in $[1\sigma, 2\sigma]$ level, yellow: in $[2\sigma, 3\sigma]$ level, red: in $[3\sigma, 5 \sigma]$ level. The color legends are the same in the figures afterwards.}
 \label{fig:i-nh}
\end{figure}

In Fig.~\ref{fig:i-nh}, we show the allowed region in case of $\tau=i$, where the left-top panel is the one for real and imaginary parts
of $\tau$,
the right-top one is the correlation between Dirac CP and Majorana phases, and the left-bottom one is the correlation between the masses of total neutrinos and neutrinoless double beta decay.
 Each of color represents the $\chi^2$ result corresponding to the level of fitting as follows; blue: within $1\sigma$ level, green: in $[1\sigma, 2\sigma]$ level, yellow: in $[2\sigma, 3\sigma]$ level, red: in $[3\sigma, 5 \sigma]$ level. 
%
The left-top one implies that the allowed region of $\tau$ is a little deviated from the exact fixed point of $\tau=i$. 
The right-top one suggests that any values are allowed for both phases, even though there tend to be favored regions.
The bottom ones tell us $\sum m_i\sim$60 meV that directly comes from the experimental value, while  $\langle m_{ee}\rangle$ tends to be localized at [0--4.5] meV.


\subsubsection{$\tau=\omega$}

\begin{figure}[H]
  \includegraphics[scale=0.4]{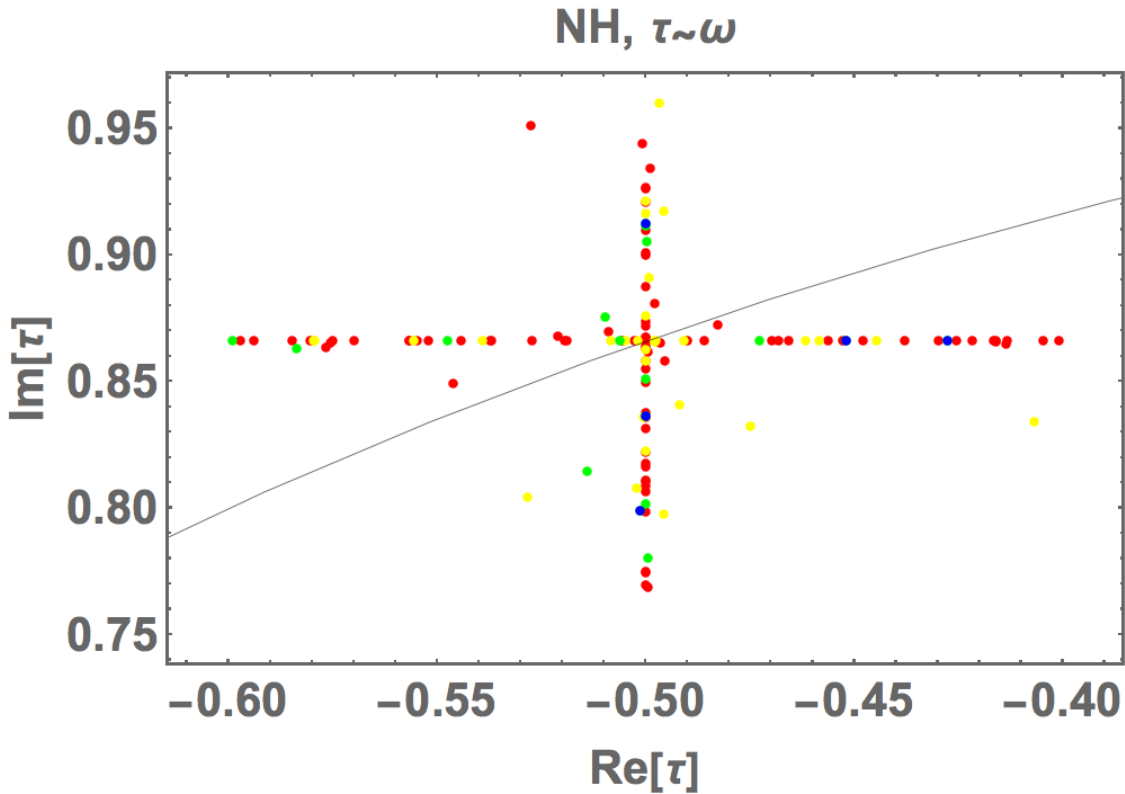}
  \includegraphics[scale=0.4]{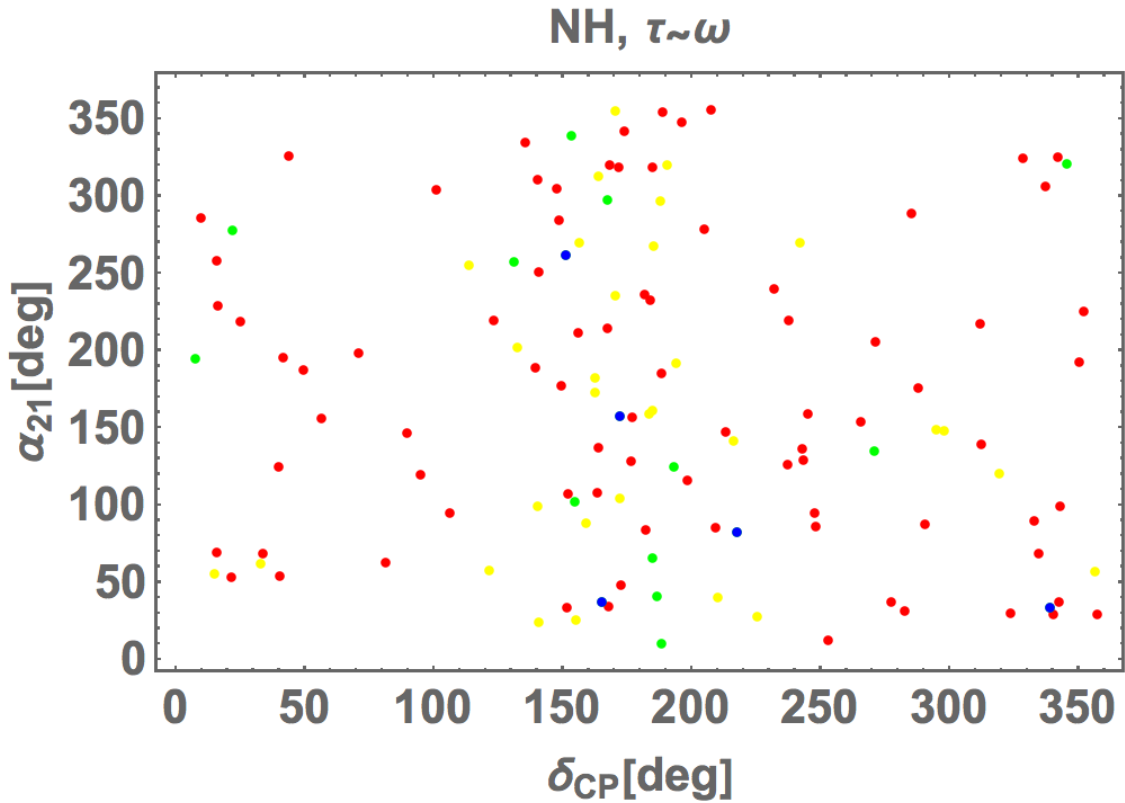}\\
   \includegraphics[scale=0.4]{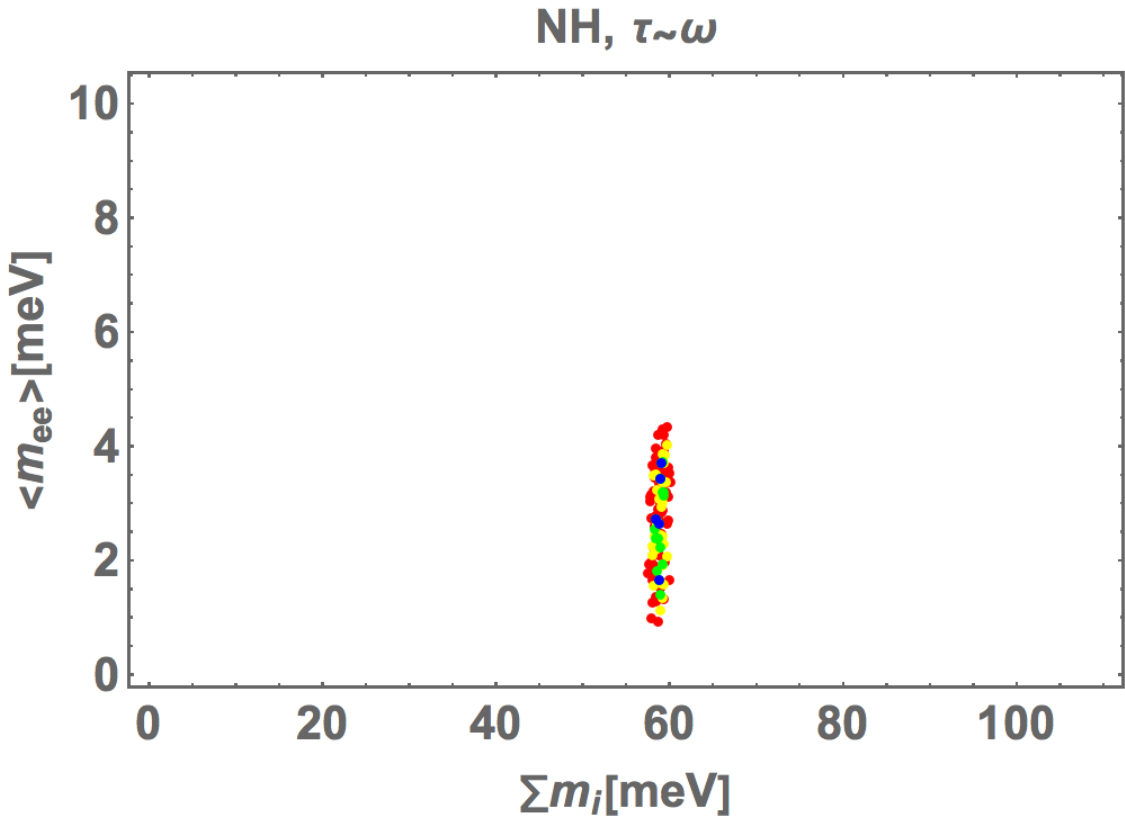}
   \includegraphics[scale=0.4]{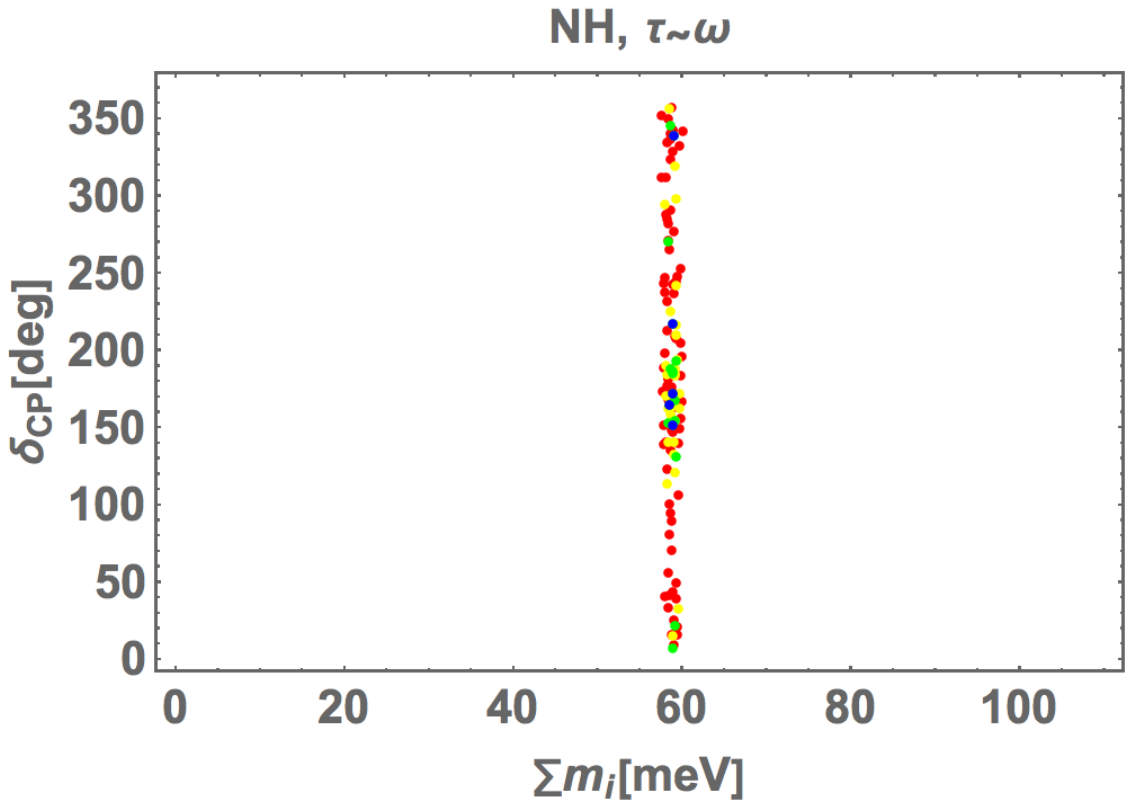}
 \caption{NH: Allowed regions in case of $\tau=\omega$. }
 \label{fig:omega-nh}
\end{figure}
In Fig.~\ref{fig:omega-nh}, we show the allowed region in case of $\tau=\omega$, where whole the legends are the same as the case of $\tau=i$. 
The left-top one implies that the allowed region of $\tau$ is at nearby $\tau=\omega$ although we would not find solutions at the exact fixed point of $\tau=\omega$. 
The right-top one suggests that any values are allowed for both phases and we do not find the correlation between them.
The bottom ones tell us $\sum m_i\sim$60 meV that directly comes from the experimental value, while  $\langle m_{ee}\rangle$ is localized at [1--4.2] meV.


\subsubsection{$\tau=-\frac12+\frac{\sqrt{15}}{2}i$}

\begin{figure}[H]
  \includegraphics[scale=0.35]{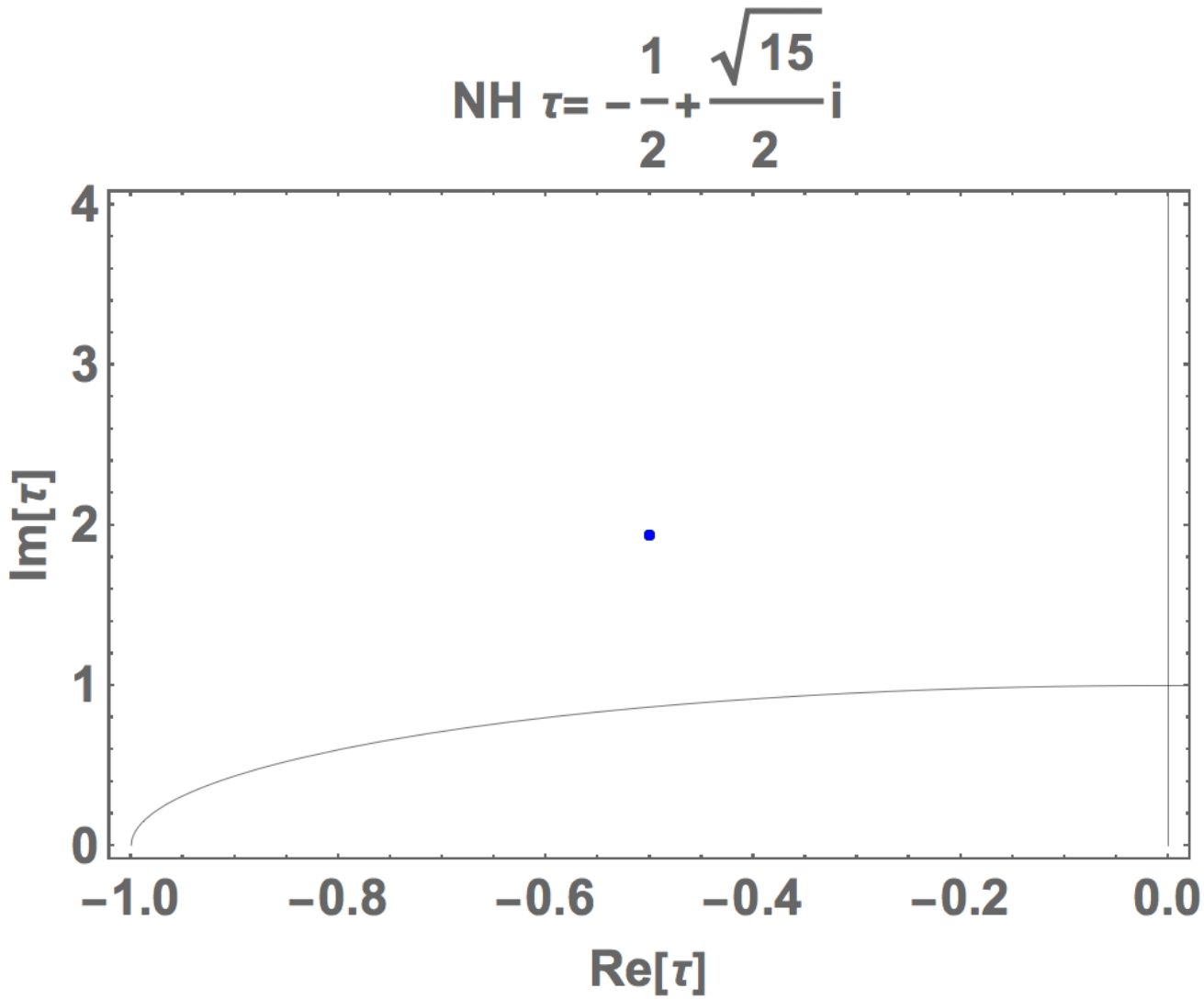}
  \includegraphics[scale=0.35]{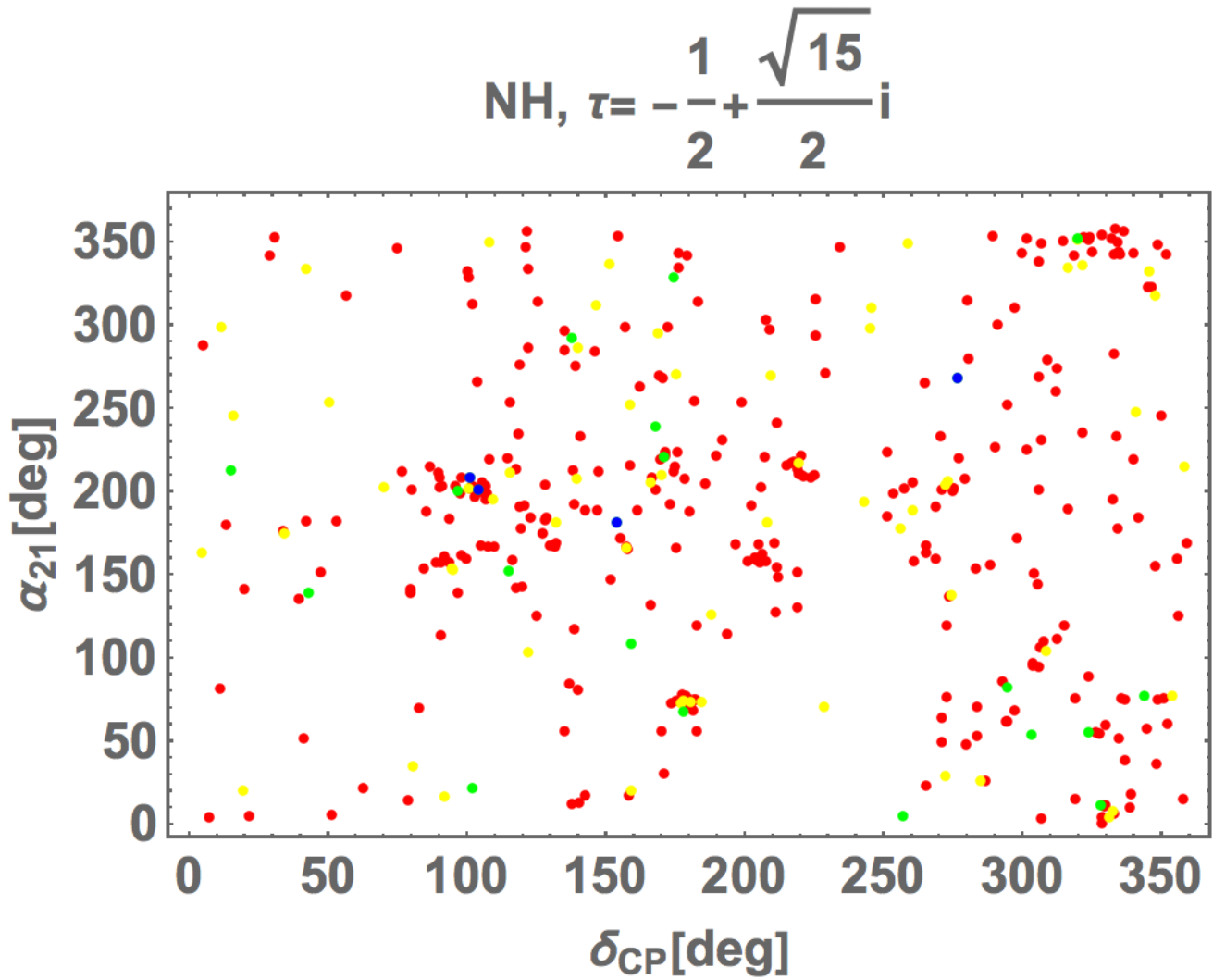}\\
   \includegraphics[scale=0.35]{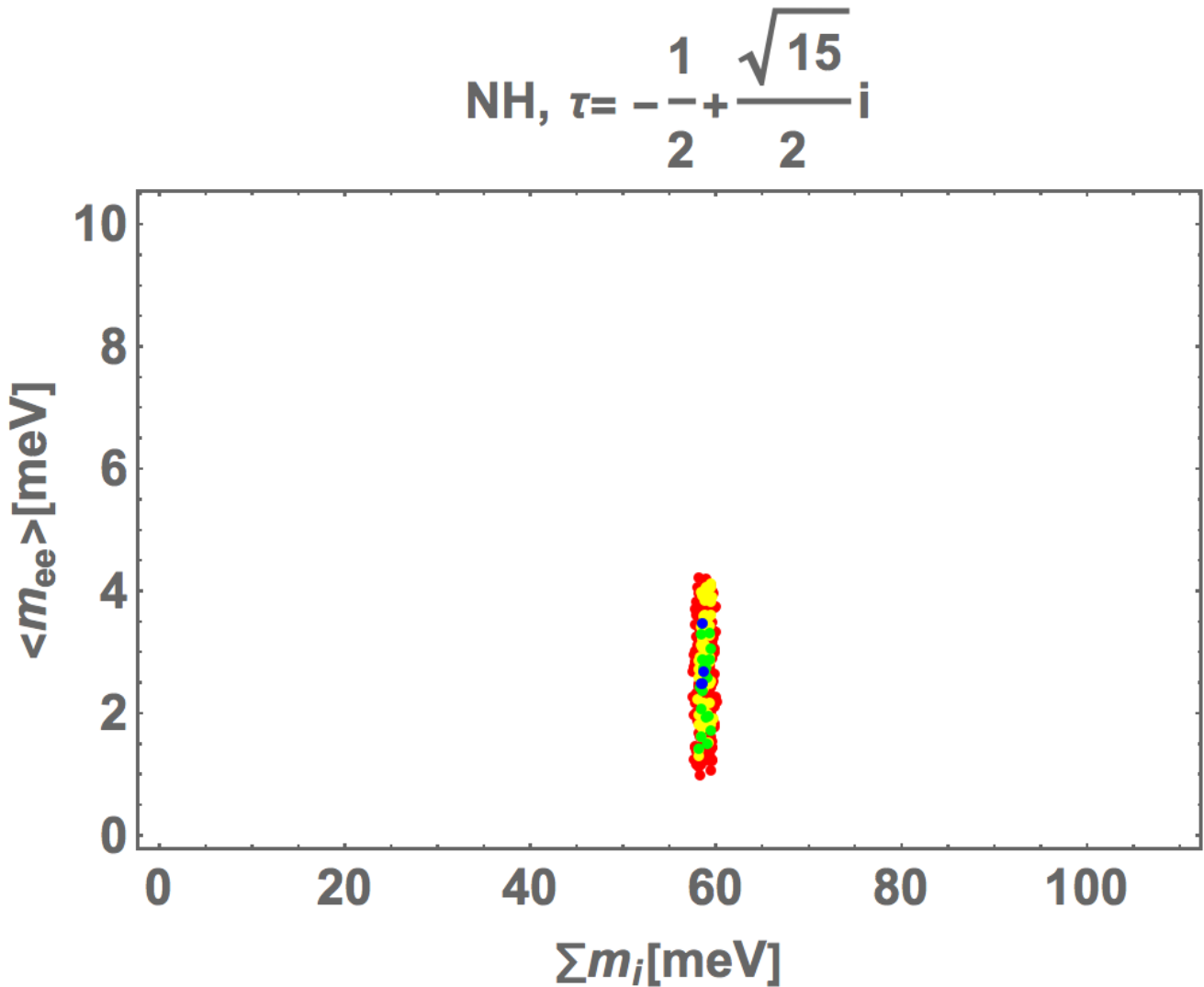}
   \includegraphics[scale=0.35]{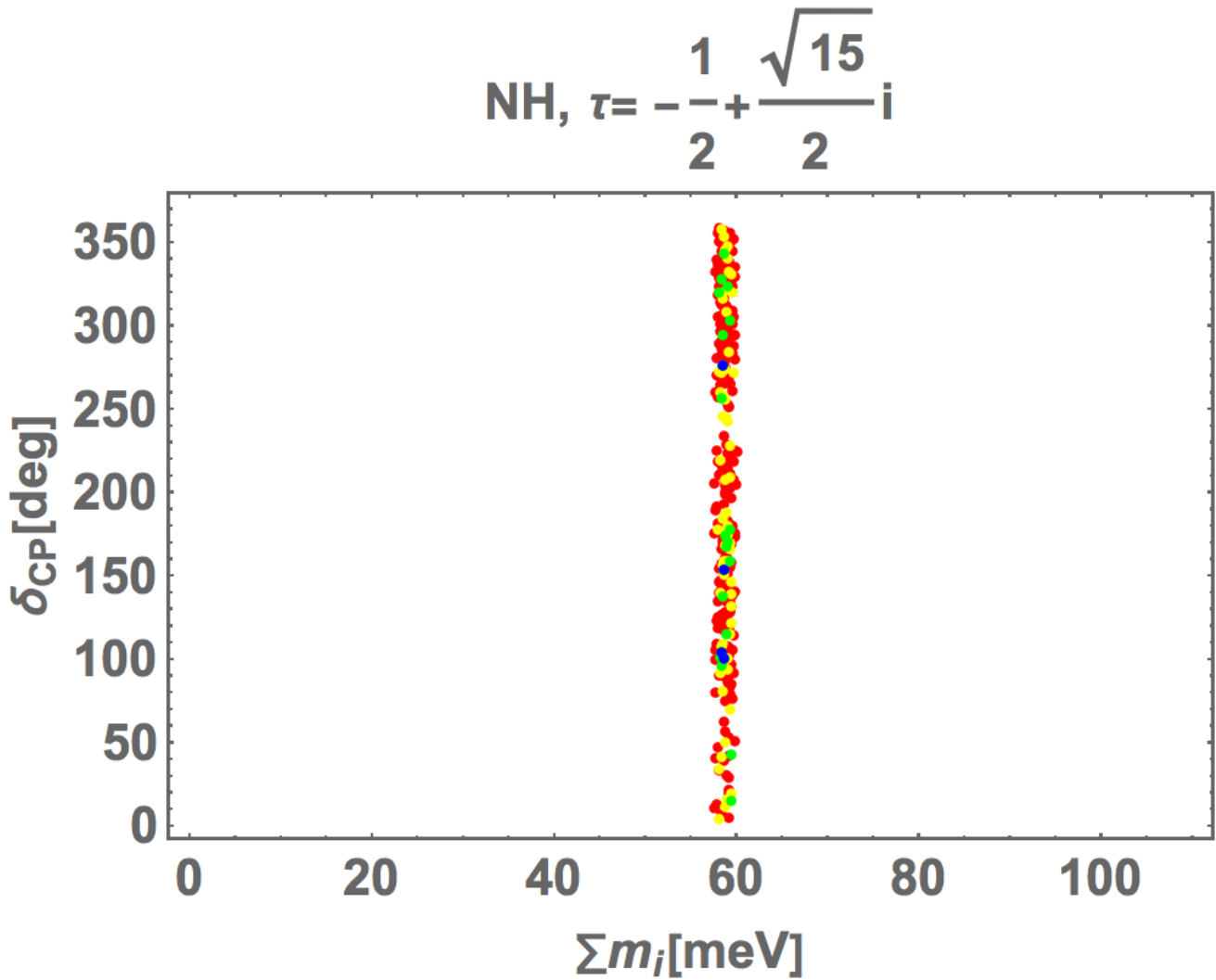}
 \caption{NH: Allowed regions in case of $\tau=-\frac12+\frac{\sqrt{15}}{2}i$.}
 \label{fig:omp-nh}
\end{figure}
In Fig.~\ref{fig:omp-nh}, we show the allowed region in case of $\tau=-\frac12+\frac{\sqrt{15}}{2}i$, where whole the legends are the same as the case of $\tau=i$. 
The left-top one implies that we find solutions on $\tau=-\frac12+\frac{\sqrt{15}}{2}i$. 
The right-top one suggests that any values are allowed for both phases, even though there tend to be favored regions.
The bottom ones tell us $\sum m_i\sim$60 meV that directly comes from the experimental value, while  $\langle m_{ee}\rangle$ is localized at [0.8--4.2] meV.


\subsubsection{$\tau=\sqrt{3}i$}

\begin{figure}[H]
  \includegraphics[scale=0.4]{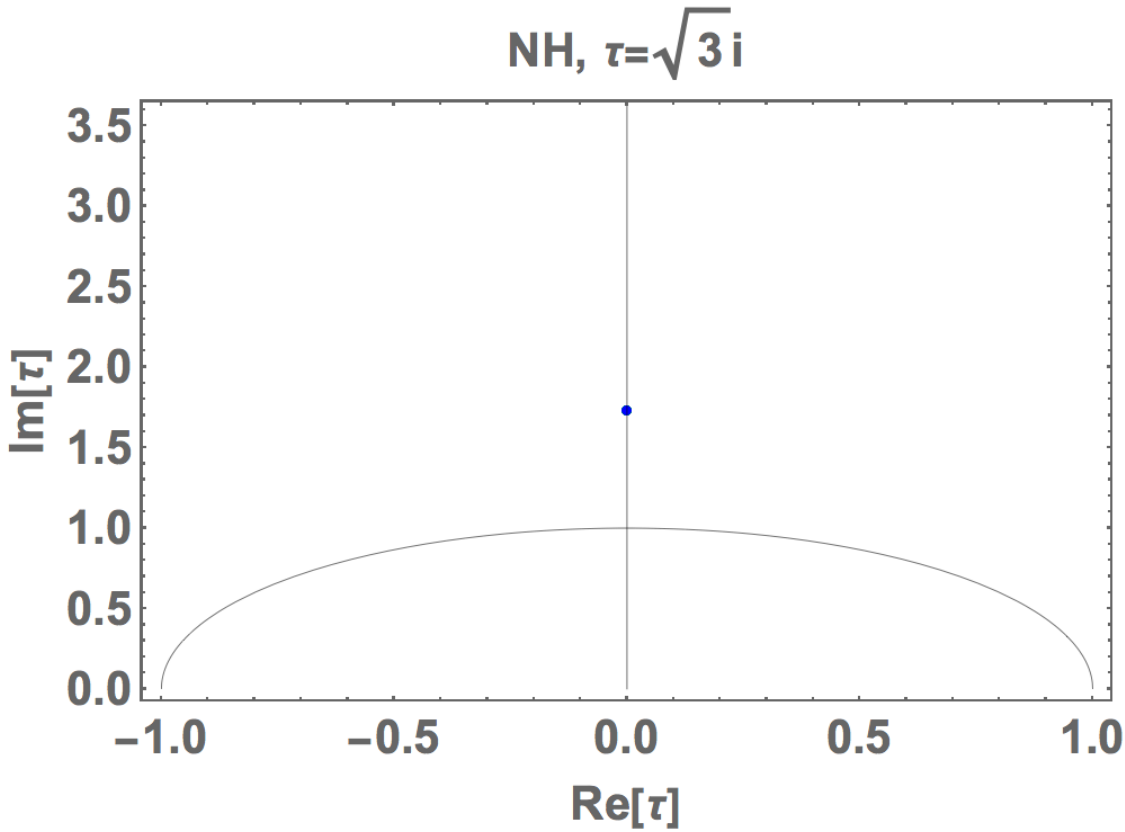}
  \includegraphics[scale=0.4]{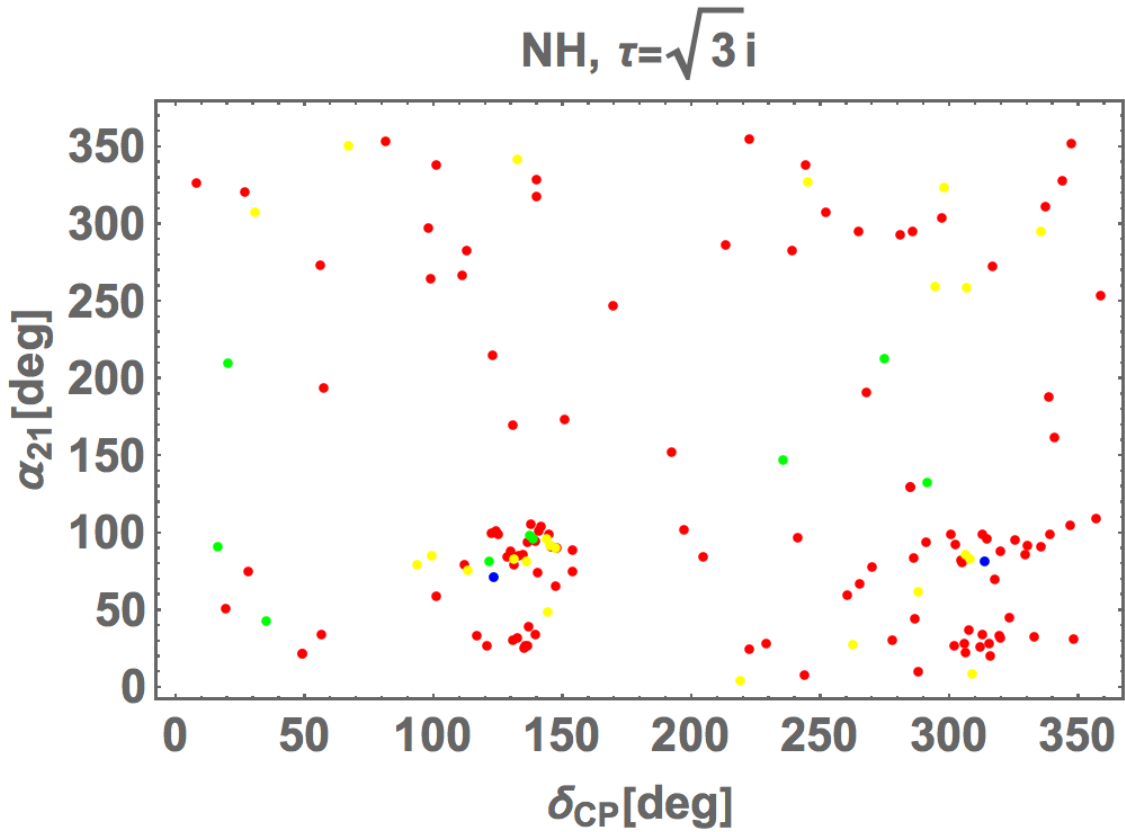}\\
   \includegraphics[scale=0.4]{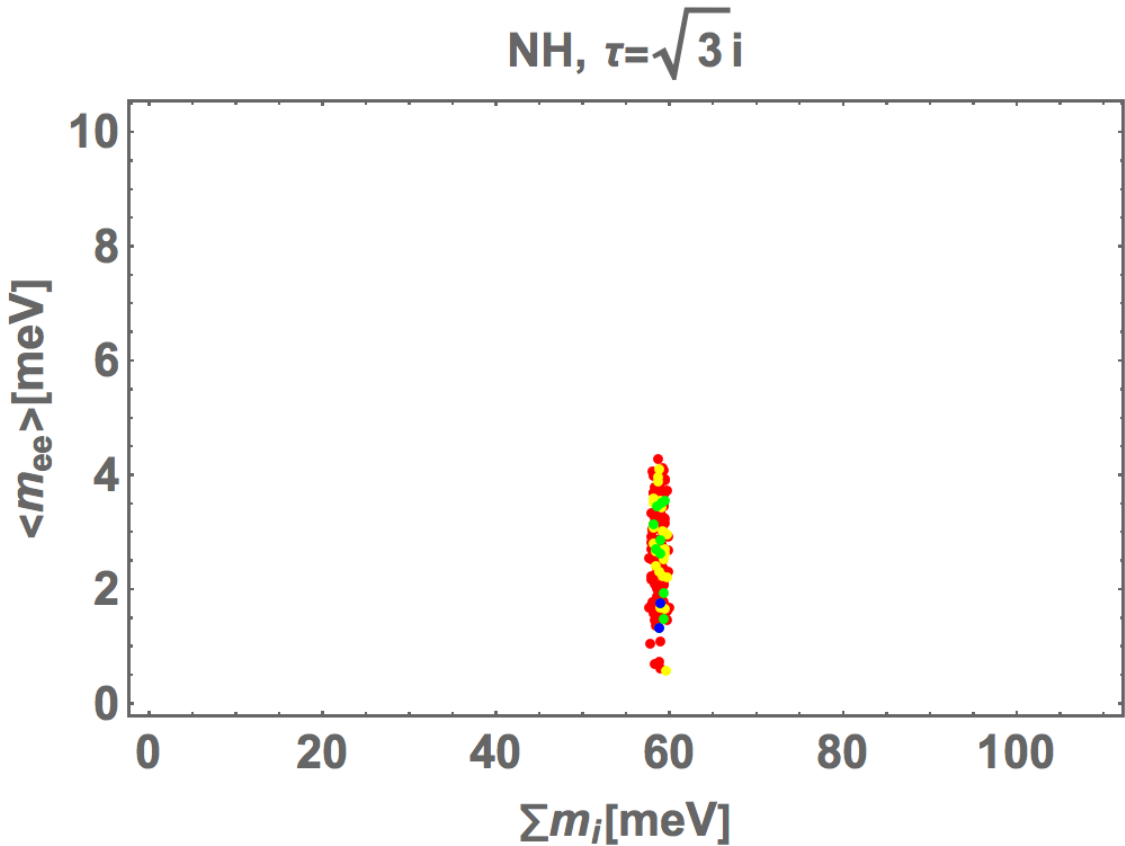}
   \includegraphics[scale=0.4]{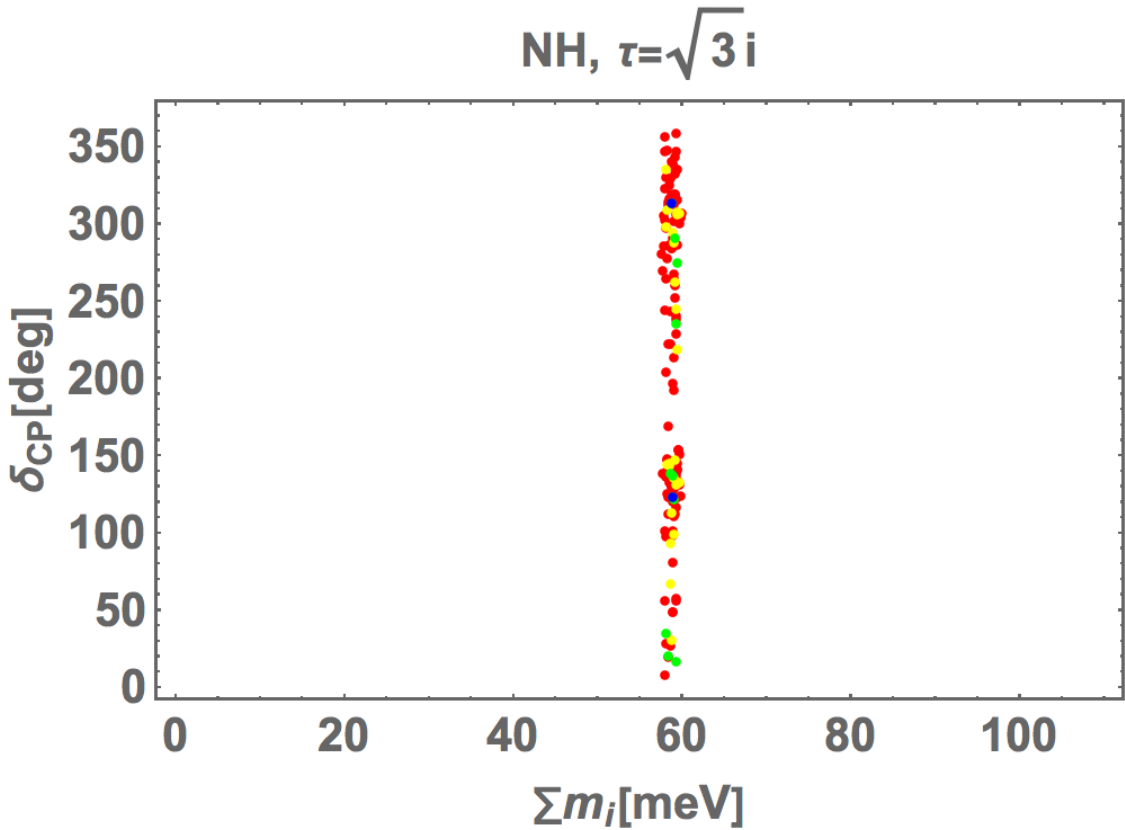}
 \caption{NH: Allowed regions in case of $\tau=\sqrt{3}i$.}
 \label{fig:r3i-nh}
\end{figure}
In Fig.~\ref{fig:r3i-nh}, we show the allowed region in case of $\tau=\sqrt{3}i$, where whole the legends are the same as the case of $\tau=i$. 
The left-top one implies that we find solutions on the exact special point $\tau=\sqrt{3}i$. 
The right-top one suggests that any values are allowed for both phases, even though there tend to be favored regions.
The bottom ones tell us $\sum m_i\sim$60 meV that directly comes from the experimental value, while  $\langle m_{ee}\rangle$ is localized at [0.5--4.2] meV.


\subsubsection{$\tau=-\frac14+\frac{\sqrt{15}}{4}i$}

\begin{figure}[H]
  \includegraphics[scale=0.4]{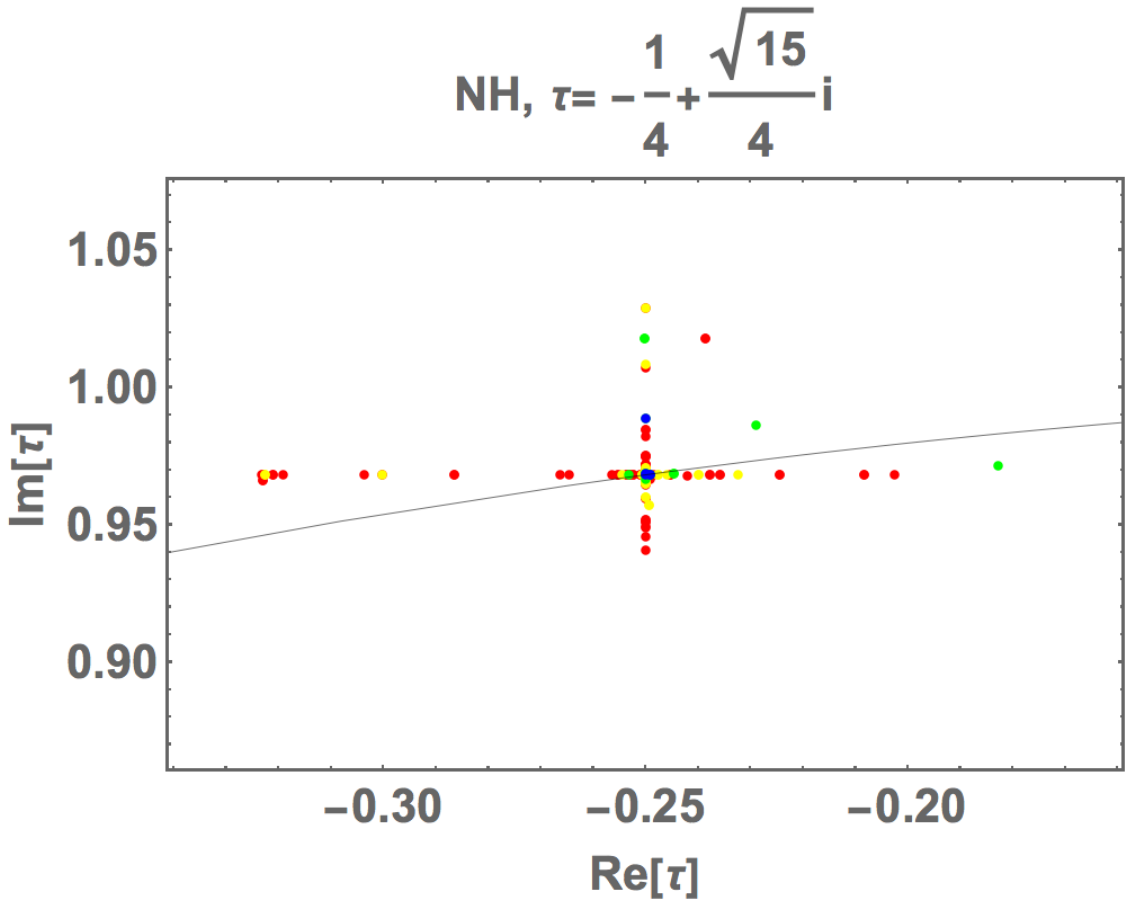}
  \includegraphics[scale=0.4]{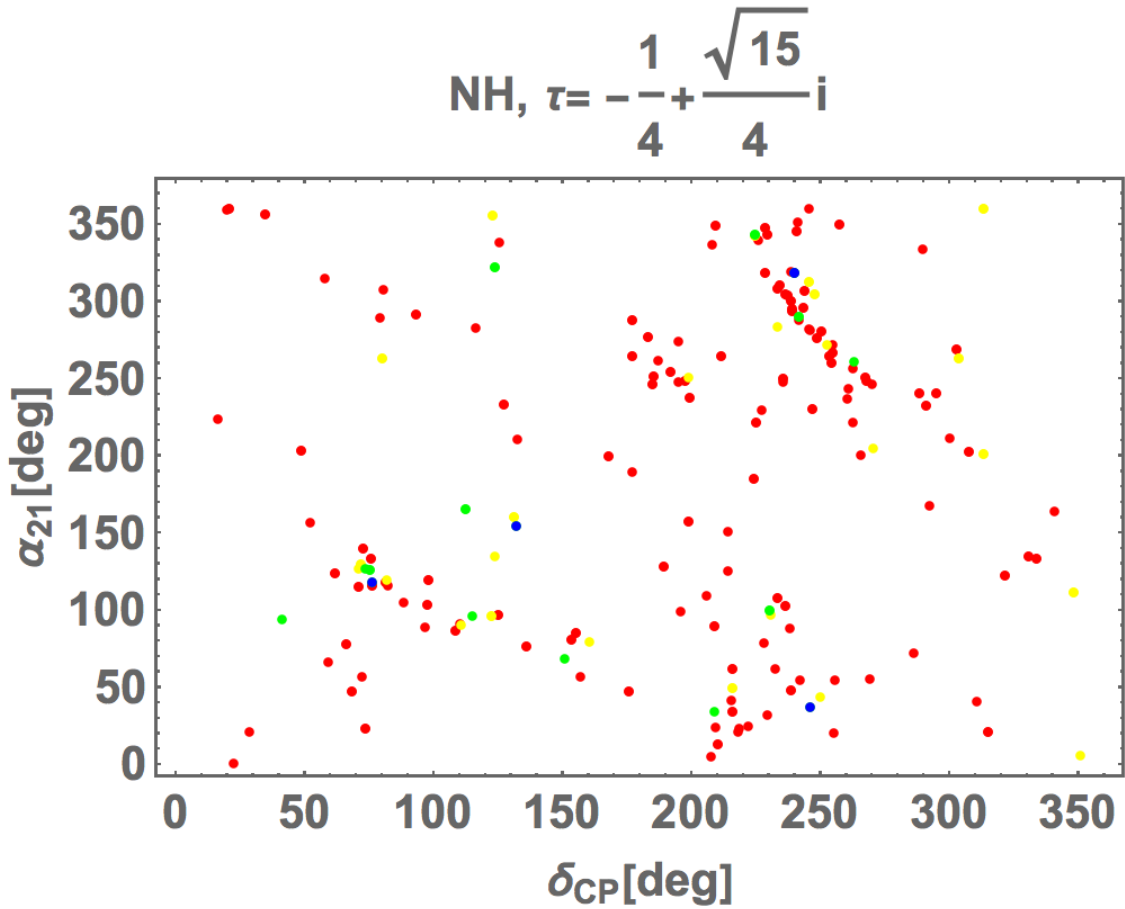}\\
   \includegraphics[scale=0.4]{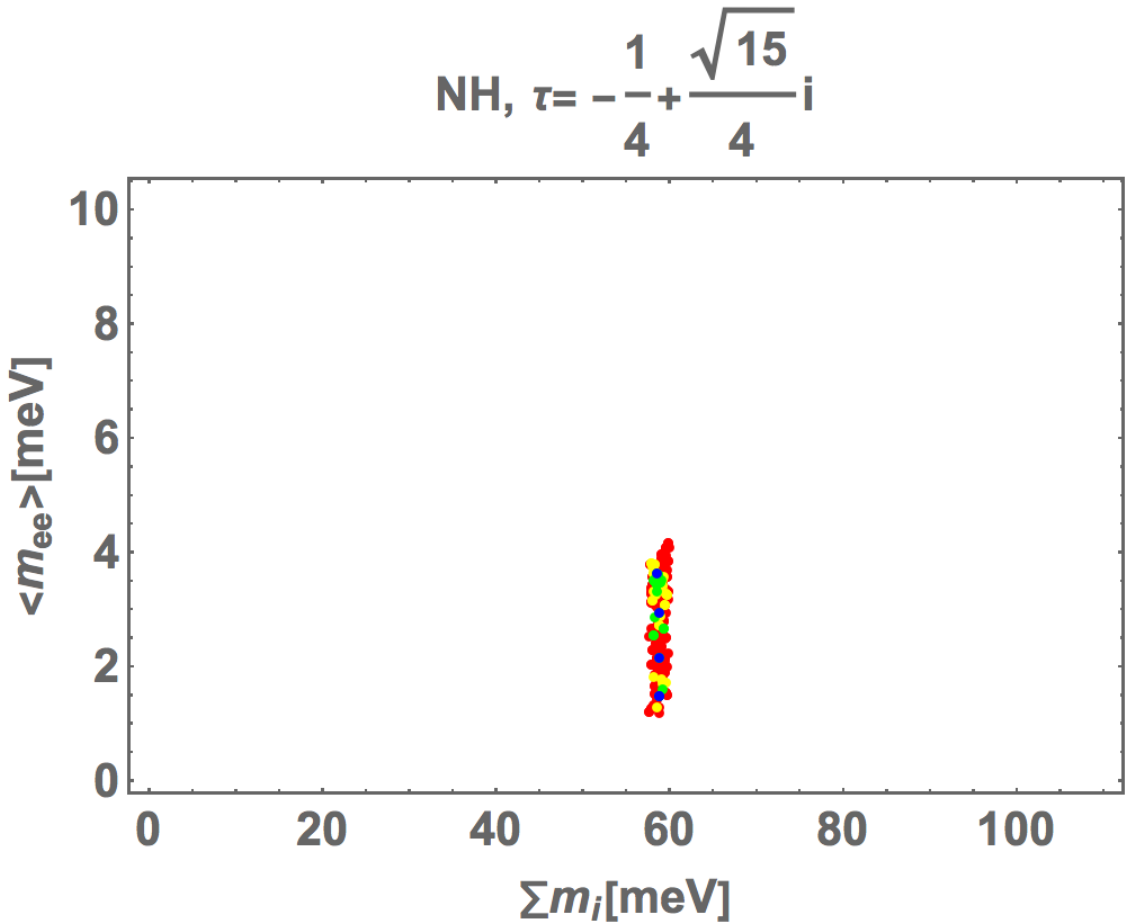}
   \includegraphics[scale=0.4]{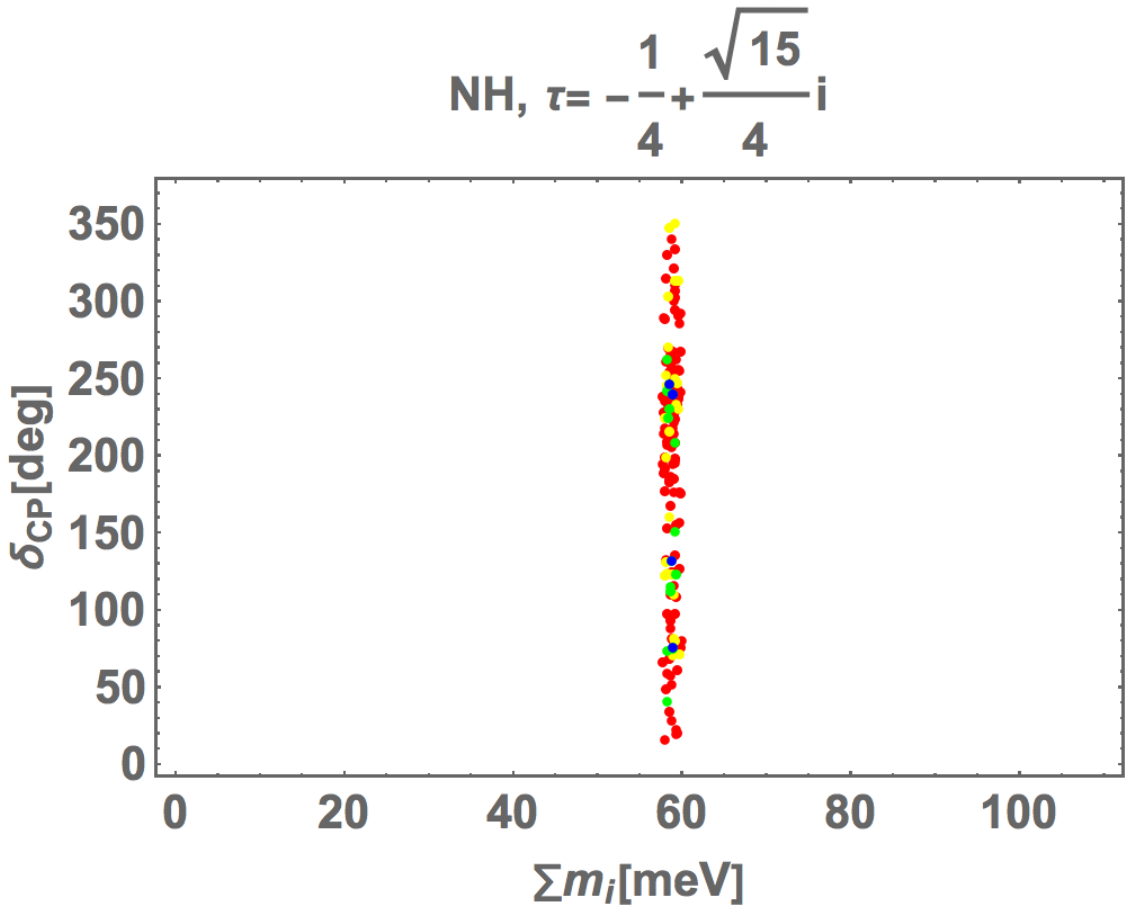}
 \caption{NH: Allowed regions in case of $\tau=-\frac14+\frac{\sqrt{15}}{4}i$.}
 \label{fig:14-nh}
\end{figure}
In Fig.~\ref{fig:14-nh}, we show the allowed region in case of $\tau=-\frac14+\frac{\sqrt{15}}{4}i$, where whole the legends are the same as the case of $\tau=i$. 
The left-top one implies that we find solutions at nearby $\tau=-\frac14+\frac{\sqrt{15}}{4}i$. 
The right-top one suggests that any values are allowed for both phases, even though there exist a few localized regions.
The bottom ones tell us $\sum m_i\sim$60 meV that directly comes from the experimental value, while  $\langle m_{ee}\rangle$ is localized at [1--4] meV.

\subsection{IH}

Here we summarize our results for the IH case at each of the fixed/special points.

\subsubsection{$\tau=i$}

\begin{figure}[H]
  \includegraphics[scale=0.35]{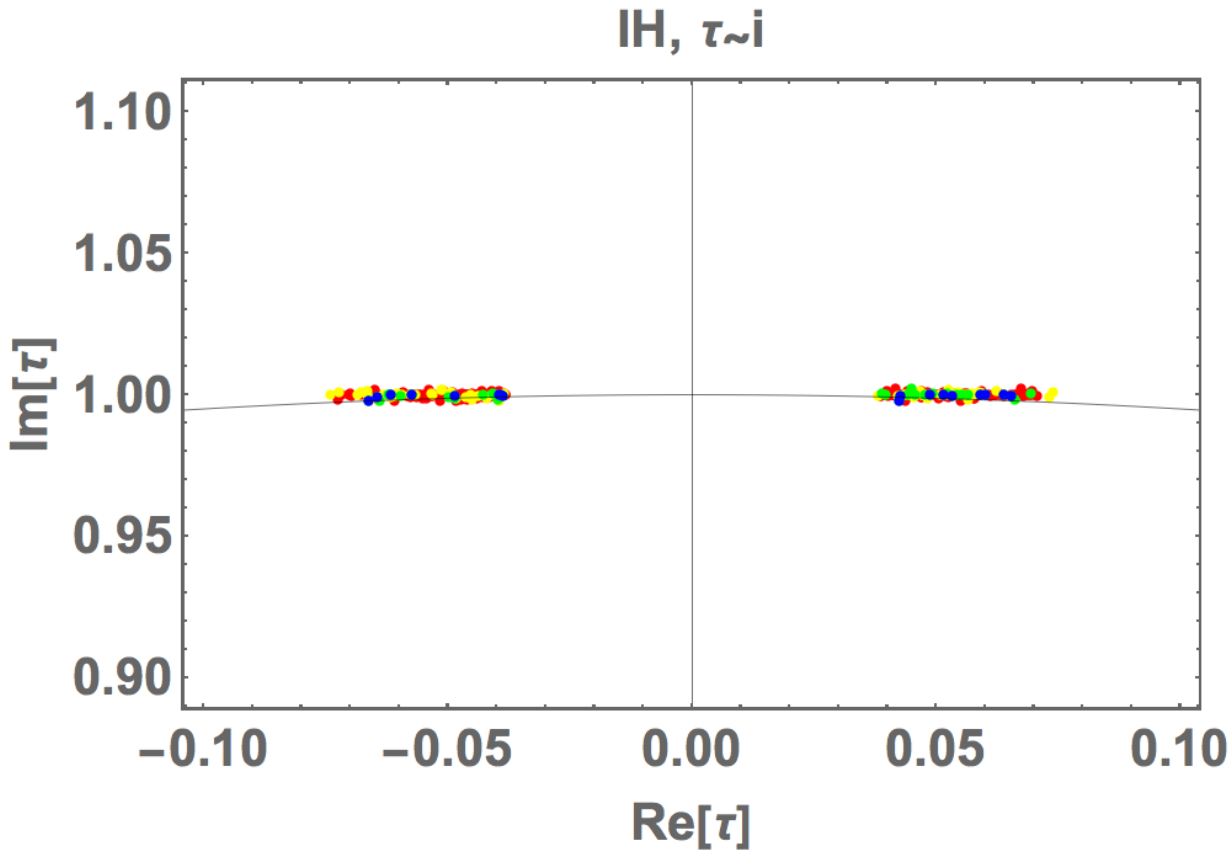}
  \includegraphics[scale=0.35]{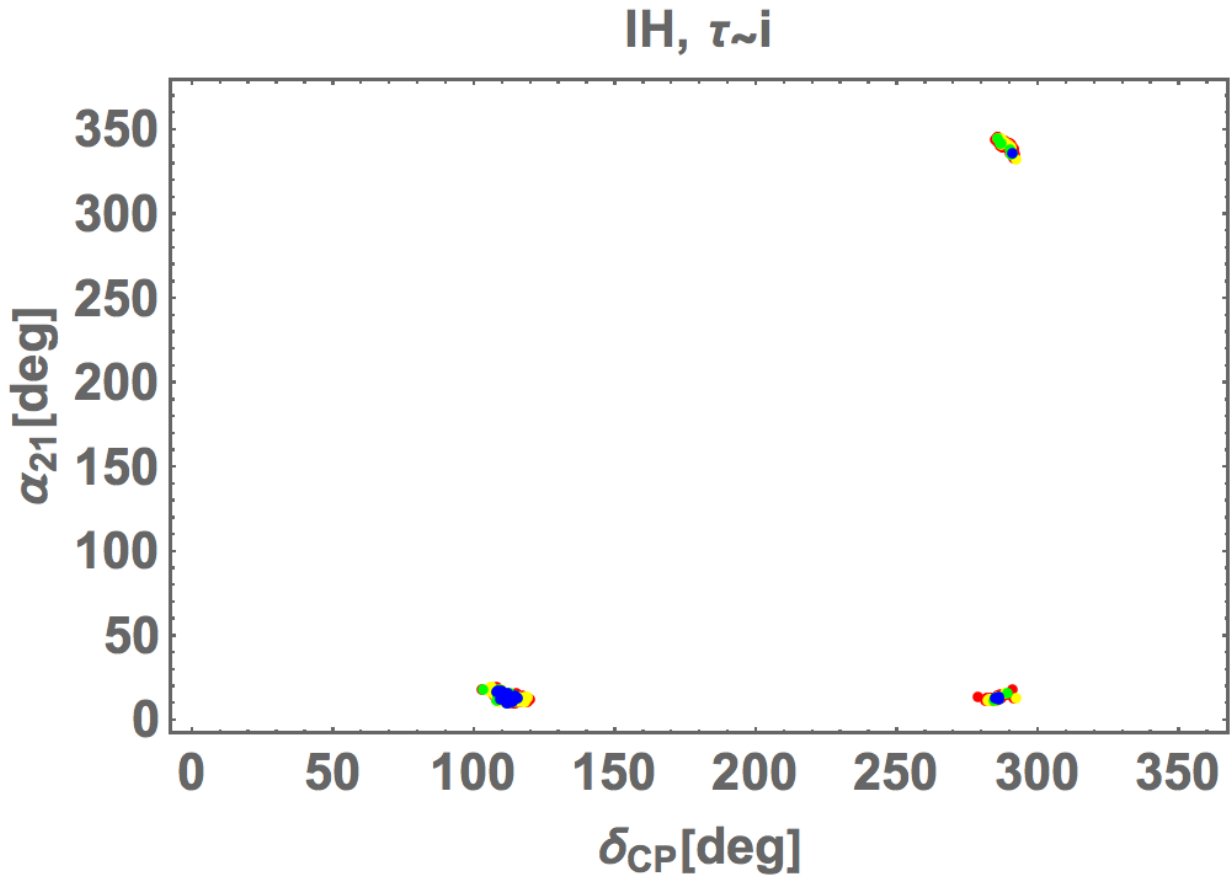}\\
   \includegraphics[scale=0.35]{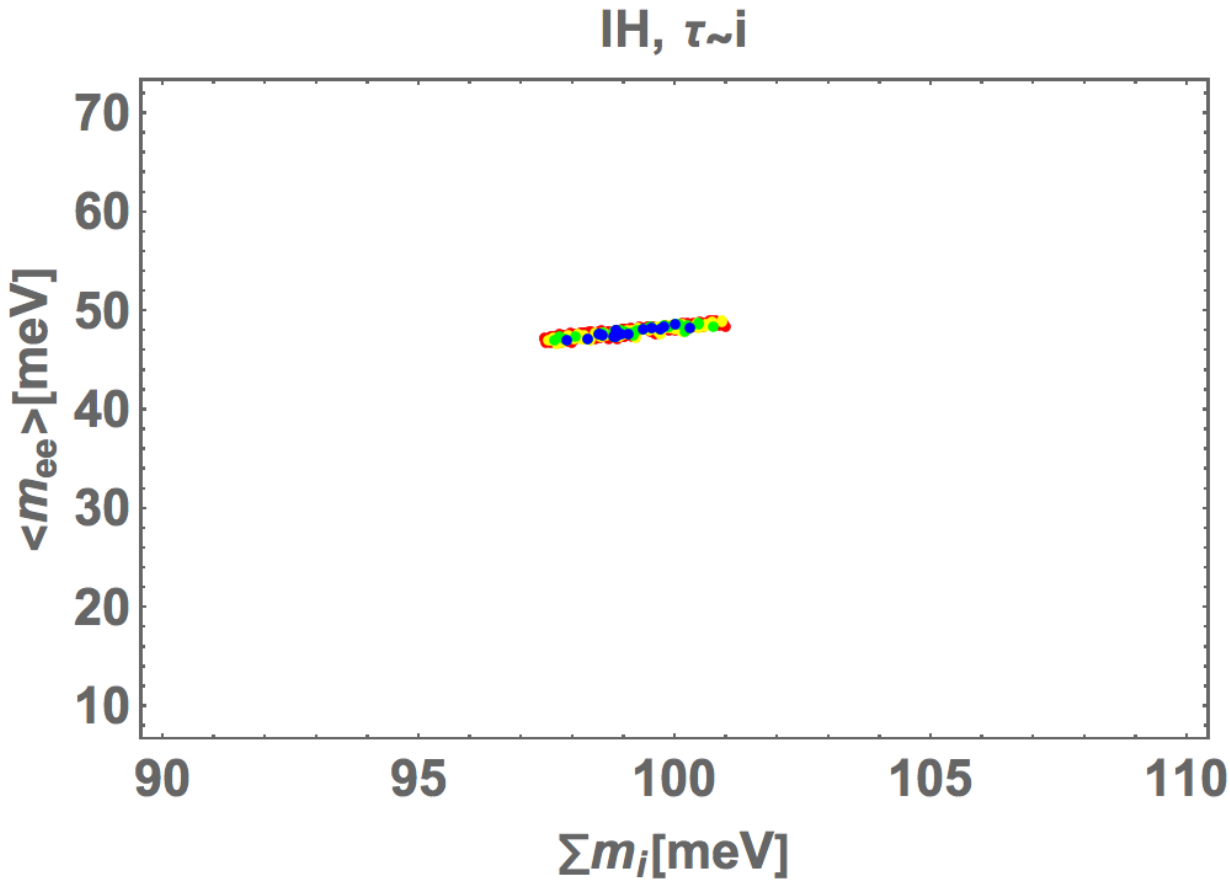}
   \includegraphics[scale=0.35]{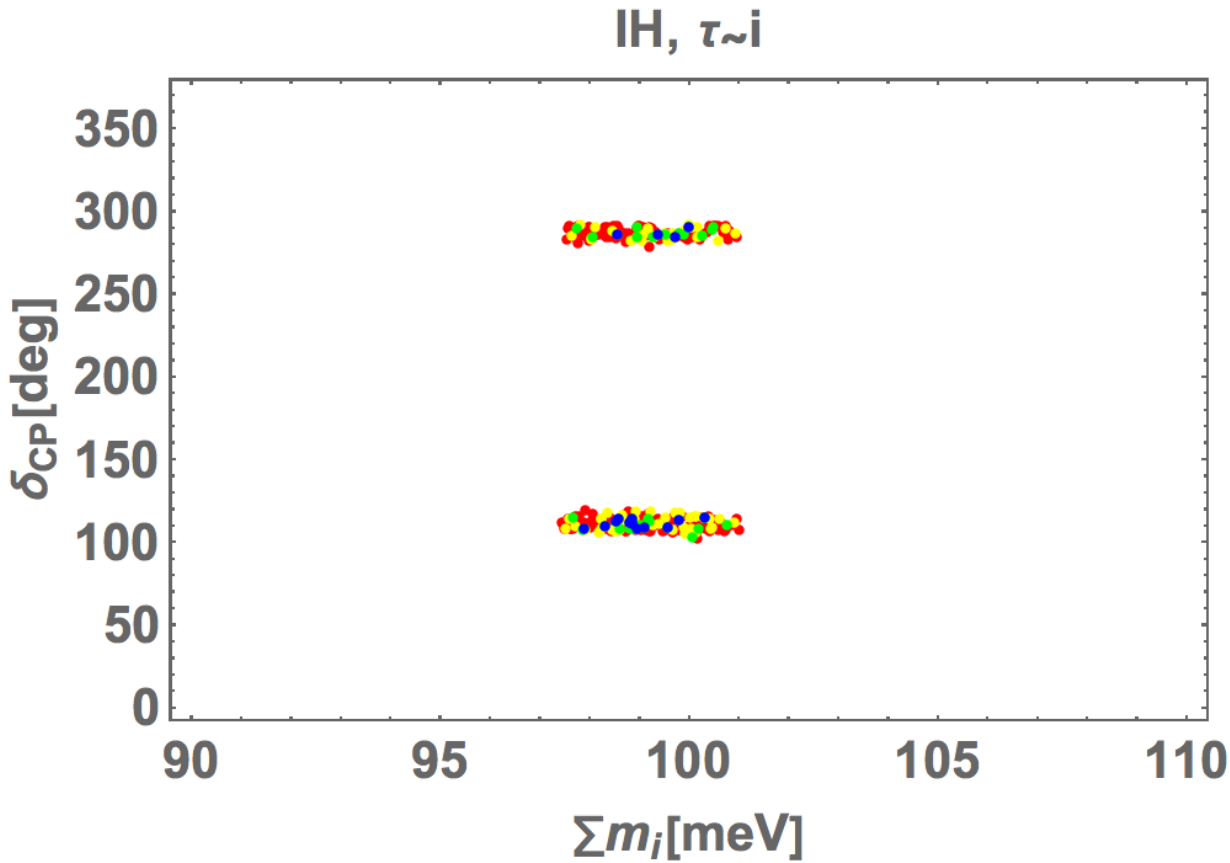}
 \caption{IH: Allowed regions in case of $\tau=i$.}
 \label{fig:i-ih}
\end{figure}
In Fig.~\ref{fig:i-ih}, we show the allowed region in case of $\tau=i$, where whole the legends are the same as the case of $\tau=i$. 
The left-top one implies that the allowed region of ${\rm Re}[\tau]$ has a little deviation from the exact fixed point of $\tau=i$. 
The right-top one suggests that  there exist three islands at around [110$^\circ$, 10$^\circ$],
[280$^\circ$, 10$^\circ$], and [280$^\circ$, 340$^\circ$] in terms of [$\delta_{\rm CP}$,$\alpha_{21}$].
The bottom ones tell us $\sum m_i=97-101$ meV that directly comes from the experimental value, while  $\langle m_{ee}\rangle$ tends to be localized at 48 meV.

\subsubsection{$\tau=\omega$}

\begin{figure}[H]
  \includegraphics[scale=0.35]{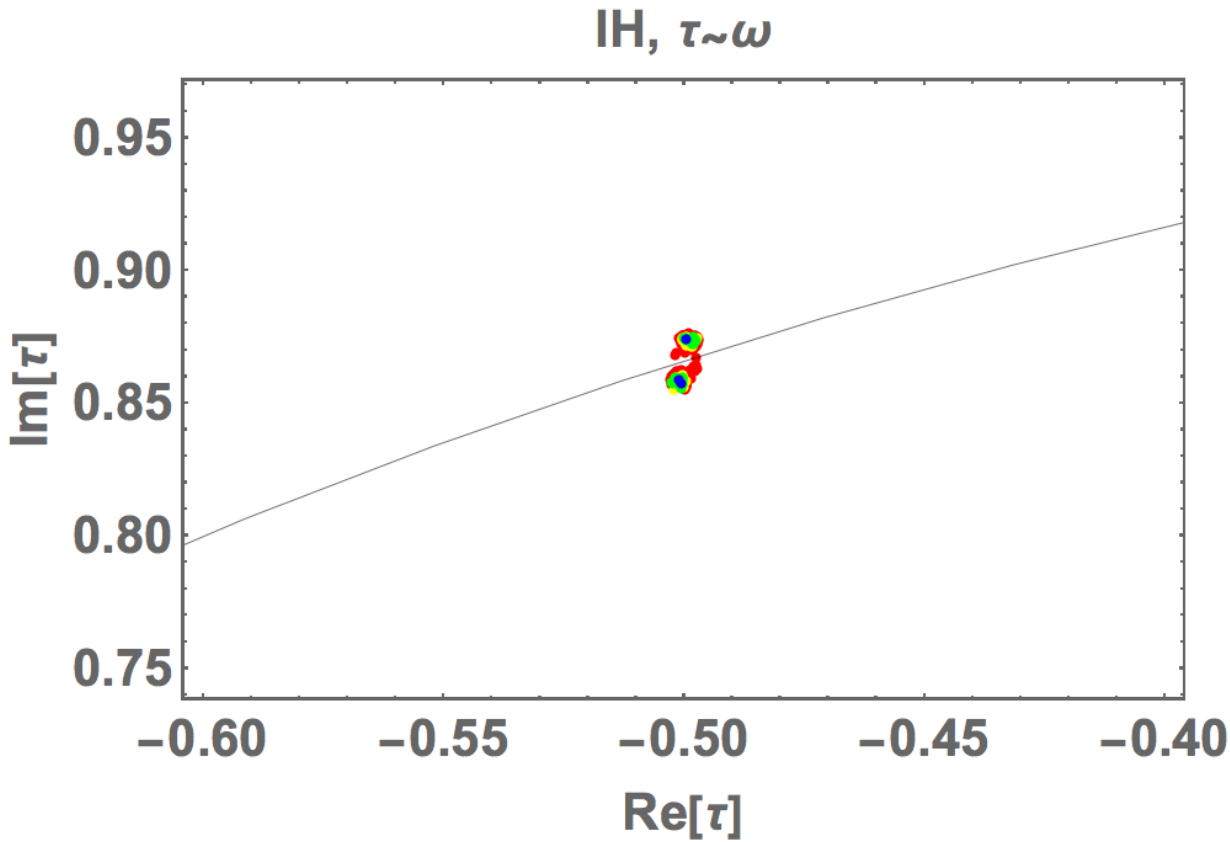}
  \includegraphics[scale=0.35]{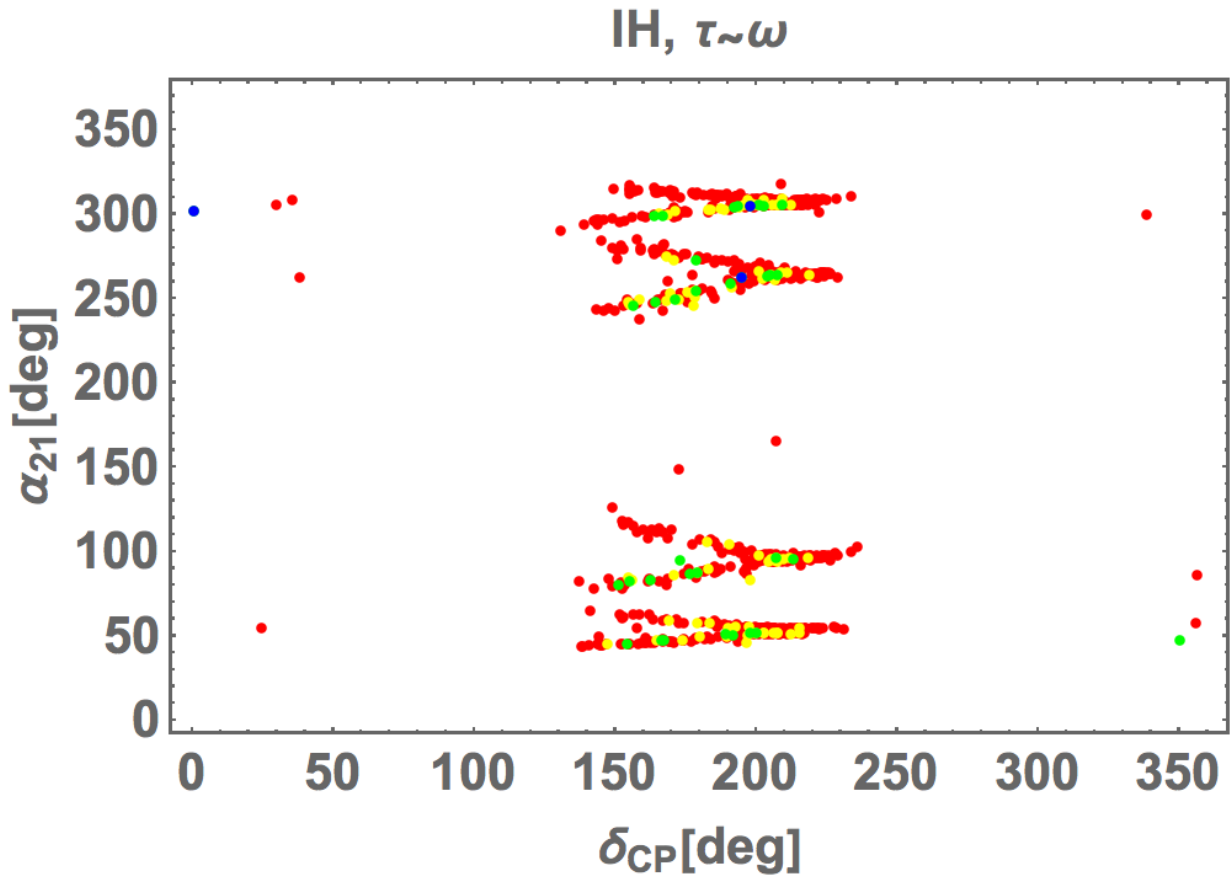}\\
   \includegraphics[scale=0.35]{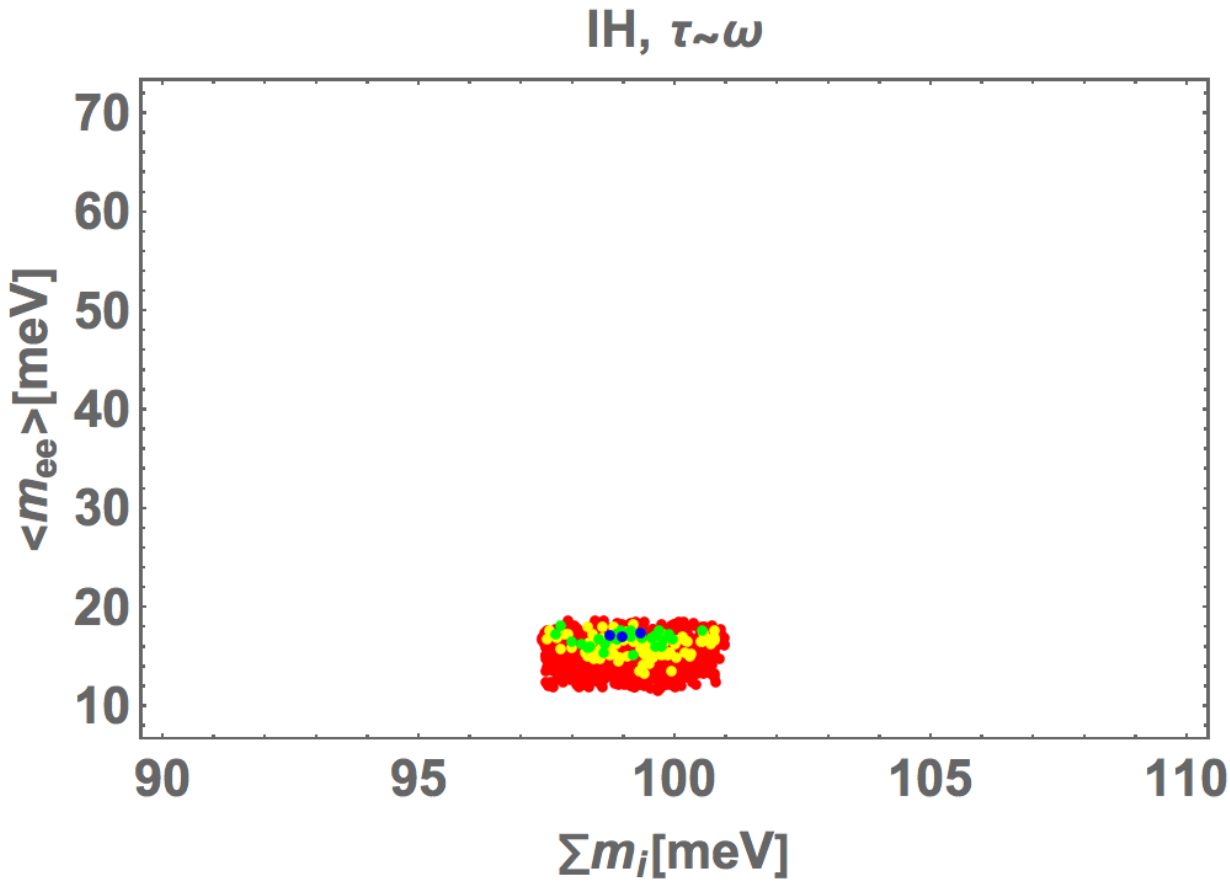}
   \includegraphics[scale=0.35]{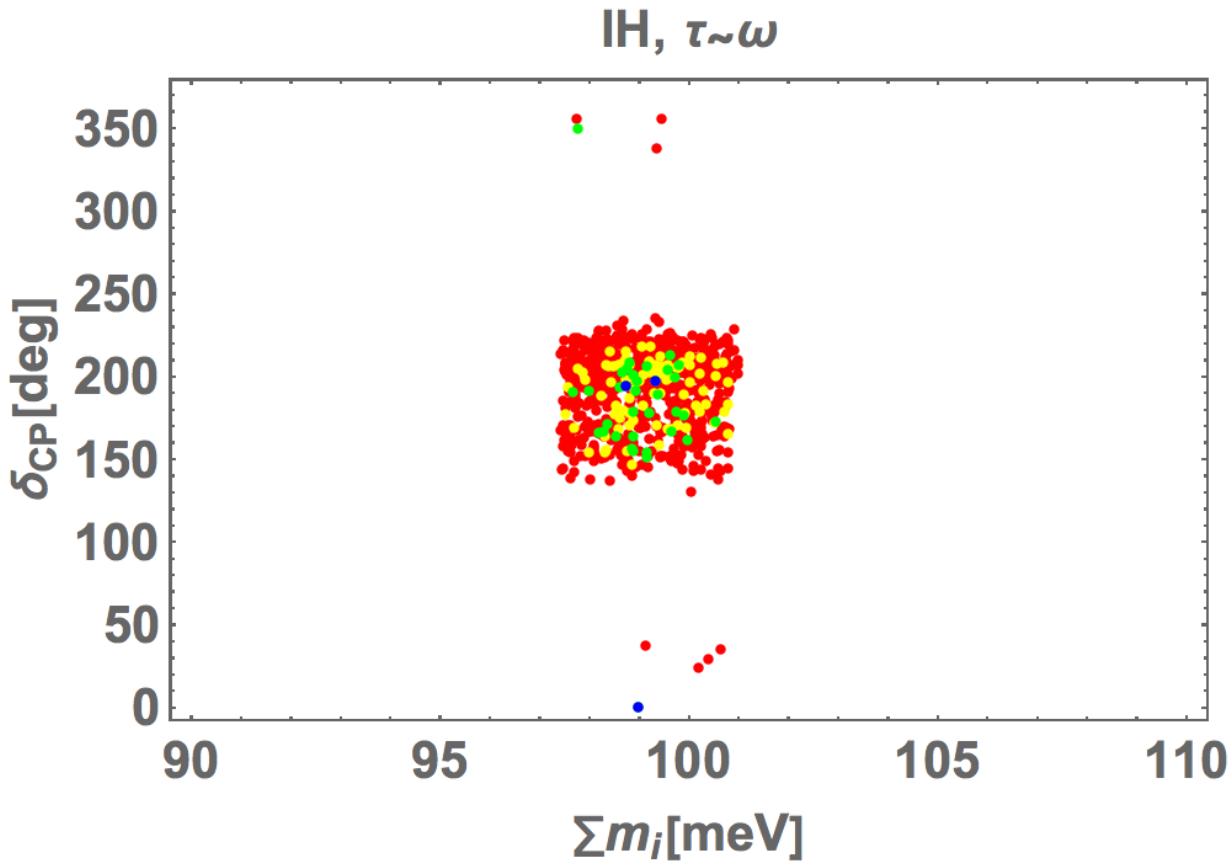}
 \caption{IH: Allowed regions in case of $\tau=\omega$.}
 \label{fig:omega-ih}
\end{figure}
In Fig.~\ref{fig:omega-ih}, we show the allowed region in case of $\tau=\omega$, where whole the legends are the same as the case of $\tau=i$. 
The left-top one implies that the allowed region of $\tau$ requires a small deviation from $\tau=\omega$. 
The right-top one suggests that  there tends to be four islands at the common region of $140^\circ\lesssim \delta_{\rm CP}\lesssim 240^\circ$, where $\alpha_{21}$ has
$40^\circ-60^\circ$, $63^\circ-130^\circ$, $240^\circ-280^\circ$, and $281^\circ-310^\circ$.
The bottom ones tell us $\sum m_i=$97-101 meV that directly comes from the experimental value, while  $\langle m_{ee}\rangle$ tends to be localized at 11-16 meV.

\subsubsection{$\tau=-\frac12+\frac{\sqrt{15}}{2}i$}

\begin{figure}[H]
  \includegraphics[scale=0.35]{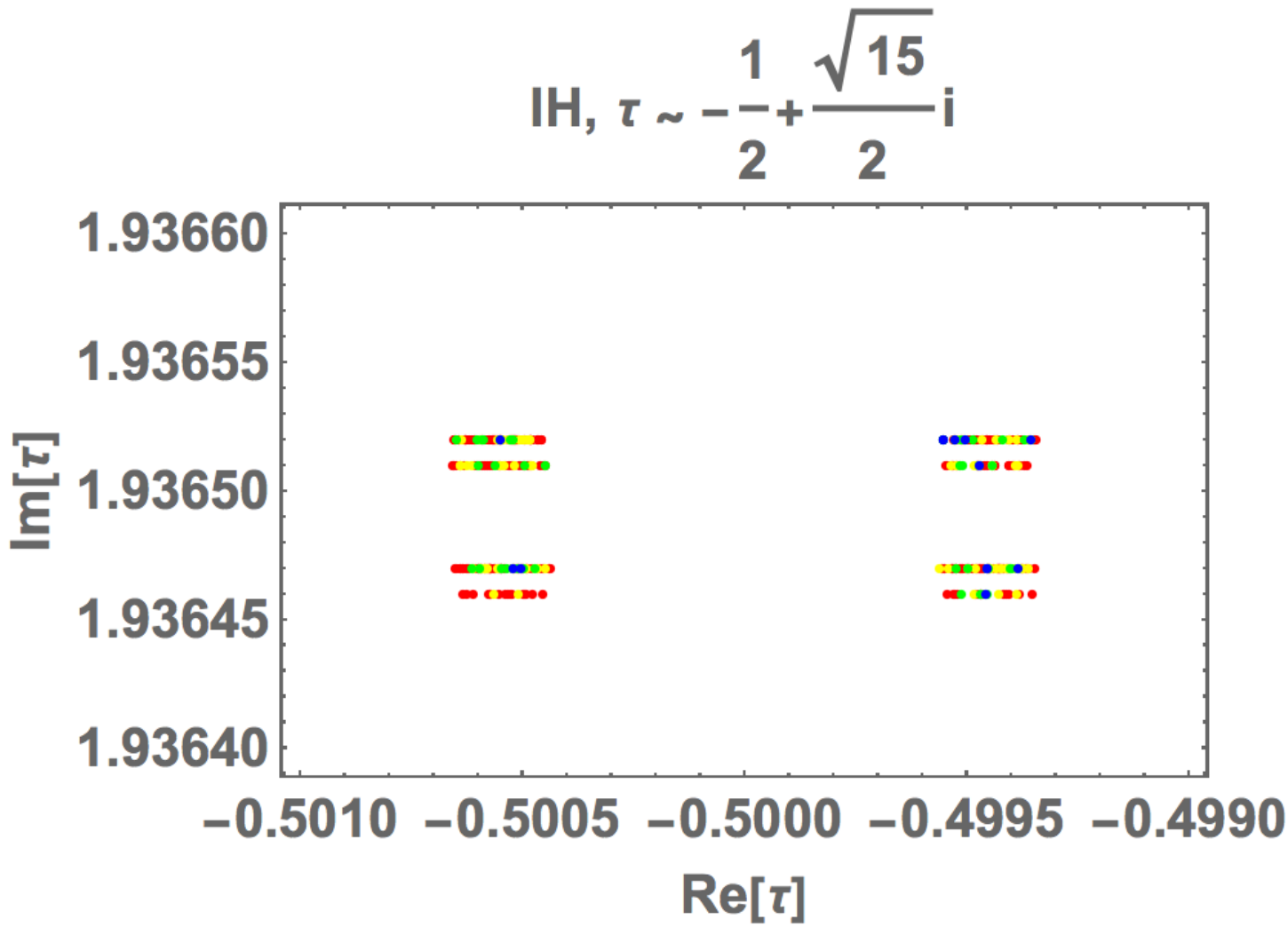}
  \includegraphics[scale=0.3]{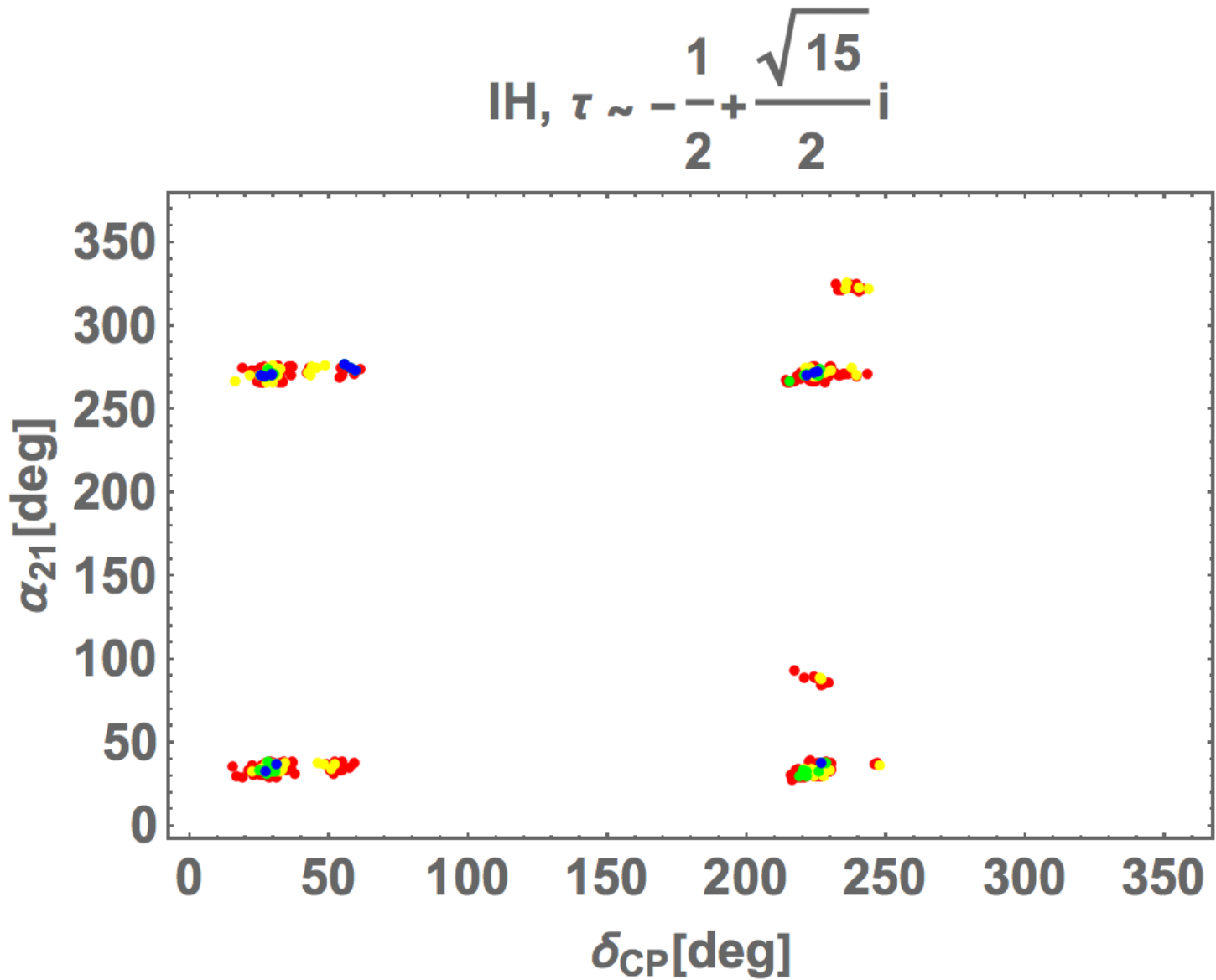}\\
   \includegraphics[scale=0.33]{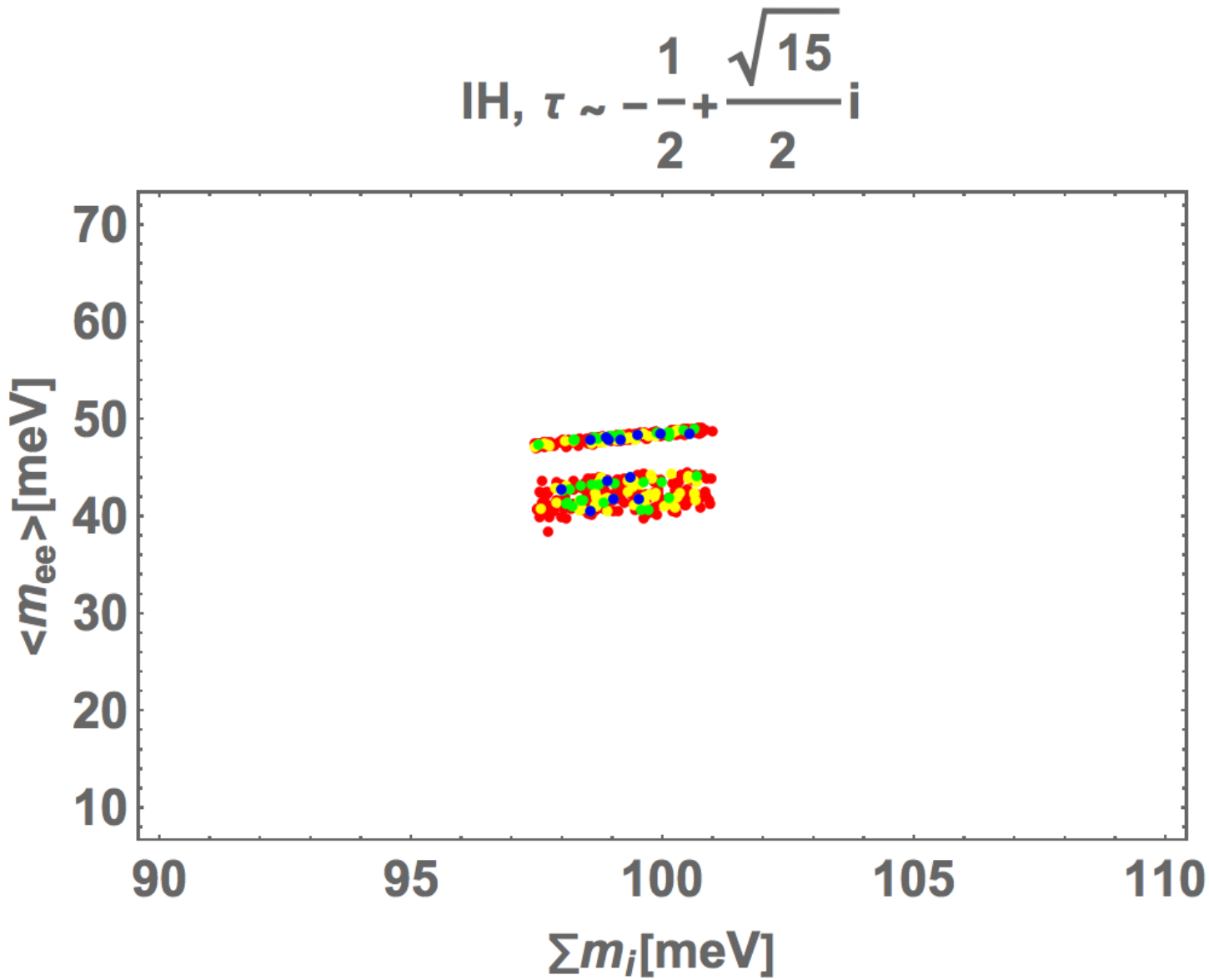}
   \includegraphics[scale=0.33]{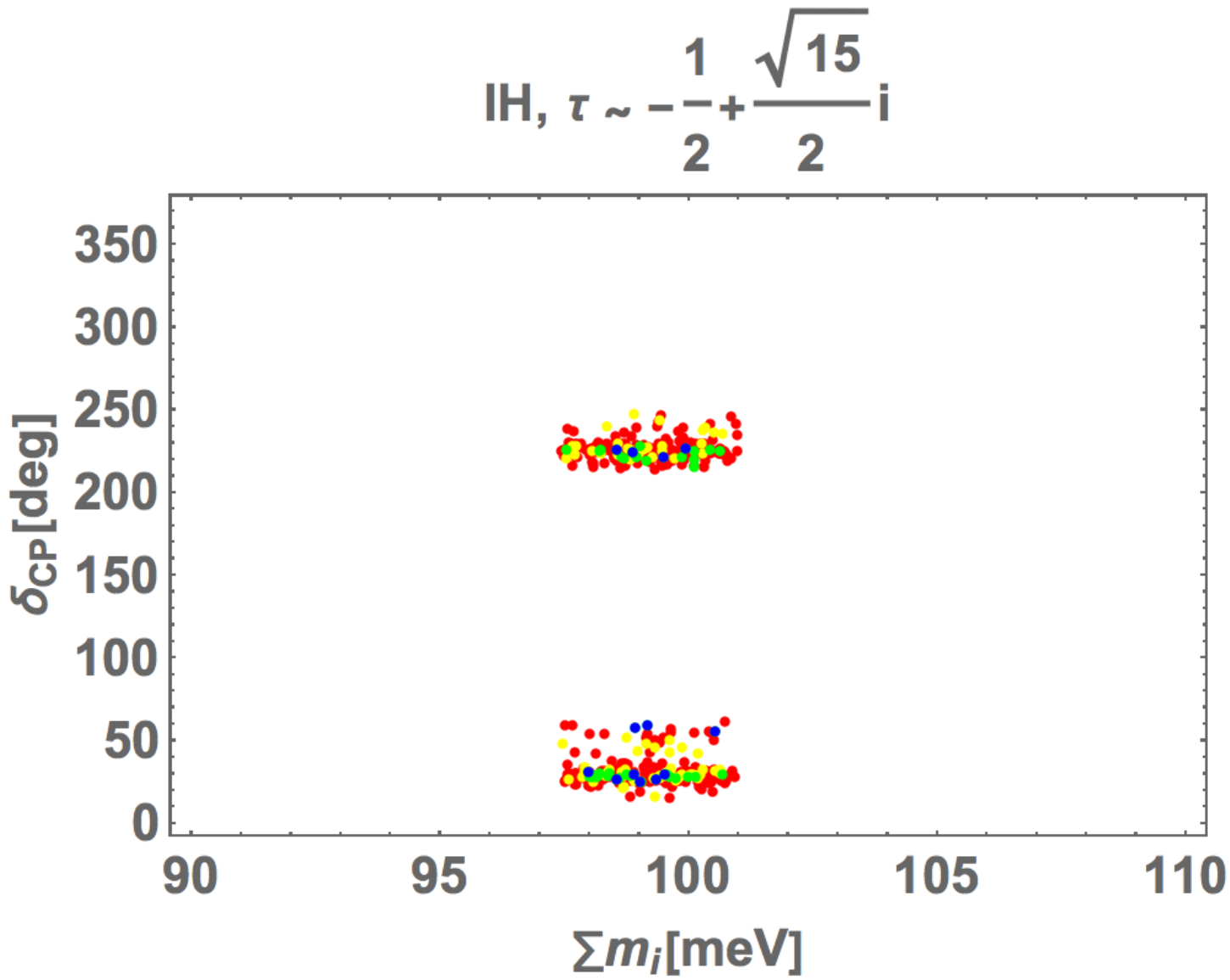}
 \caption{IH: Allowed regions in case of $\tau=-\frac12+\frac{\sqrt{15}}{2}i$.}
 \label{fig:omp-ih}
\end{figure}
In Fig.~\ref{fig:omp-ih}, we show the allowed region in case of $\tau=-\frac12+\frac{\sqrt{15}}{2}i$, where whole the legends are the same as the case of $\tau=i$. 
The left-top one implies that the allowed region of $\tau$ is divided by eight sectors and requires a small deviation from $\tau=-\frac12+\frac{\sqrt{15}}{2}i$. 
The right-top one suggests that there exist six islands; $\alpha_{21}\simeq 30^\circ, 270^\circ$  
at common region of $20^\circ\lesssim \delta_{\rm CP}\lesssim 60^\circ$, 
$\alpha_{21}\simeq 30^\circ, 80^\circ, 270^\circ, 320^\circ$  
at common region of $220^\circ\lesssim \delta_{\rm CP}\lesssim 250^\circ$.
The bottom ones tell us $\sum m_i=$97-101 meV that directly comes from the experimental value, while  $\langle m_{ee}\rangle$ tends to be localized at 39-44 and 46-50 meV.

\subsubsection{$\tau=\sqrt{3}i$}

\begin{figure}[H]
  \includegraphics[scale=0.35]{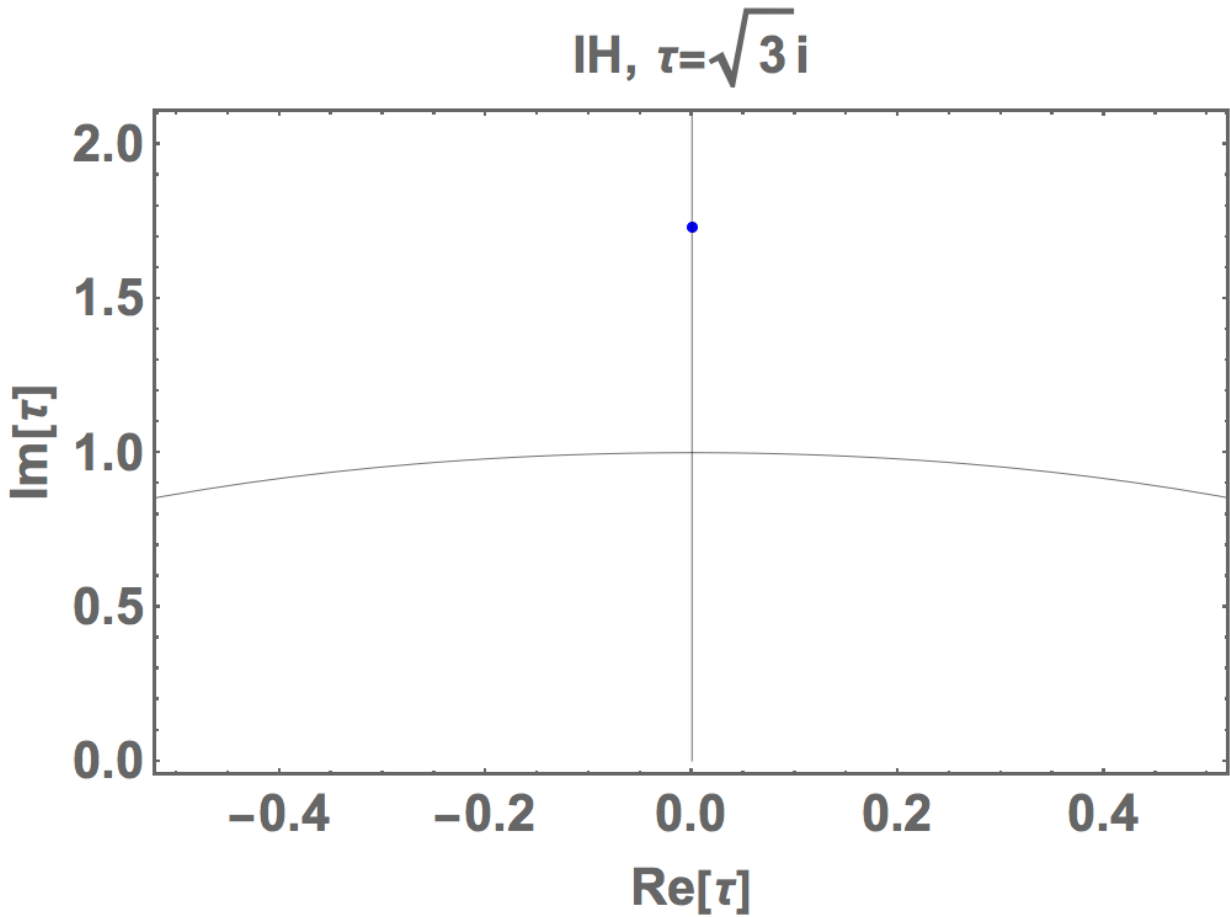}
  \includegraphics[scale=0.35]{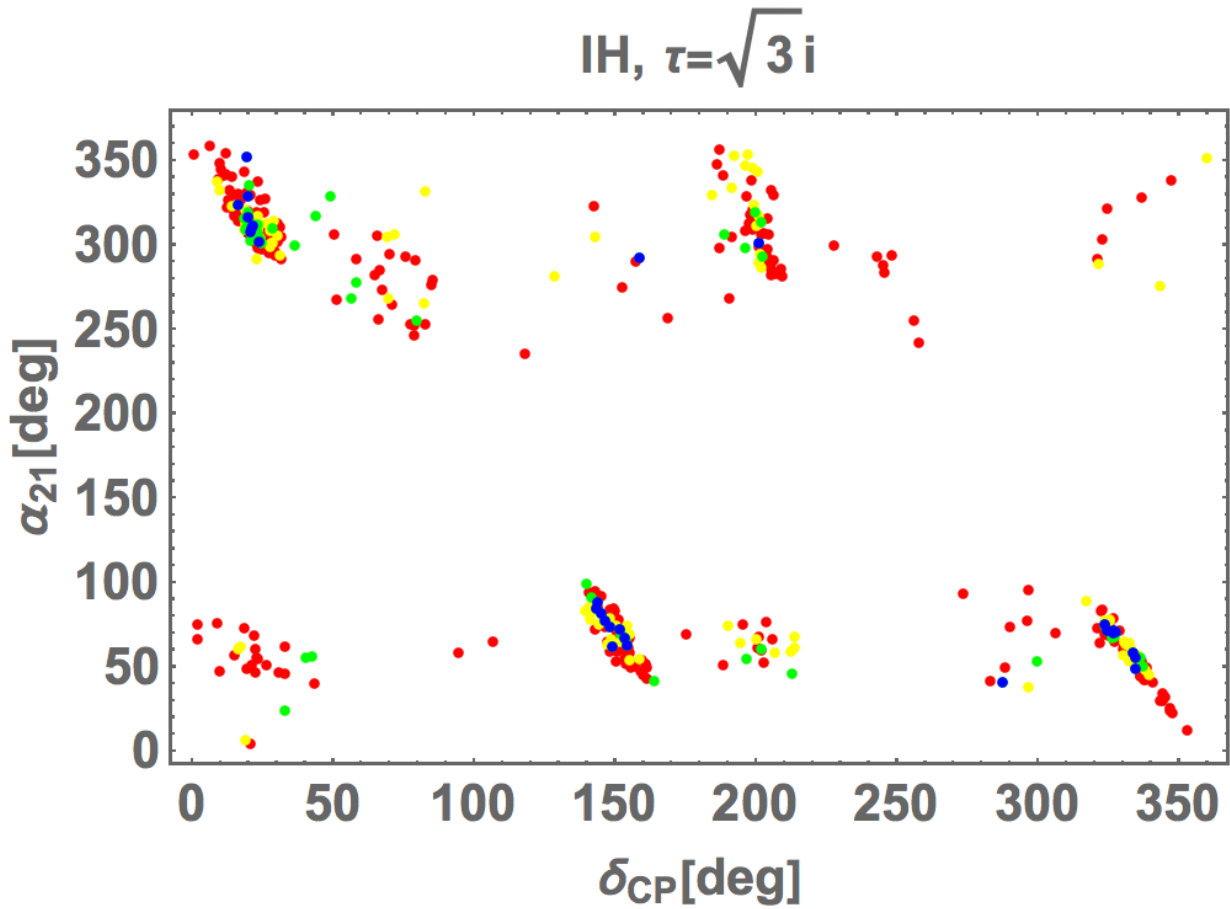}\\
   \includegraphics[scale=0.35]{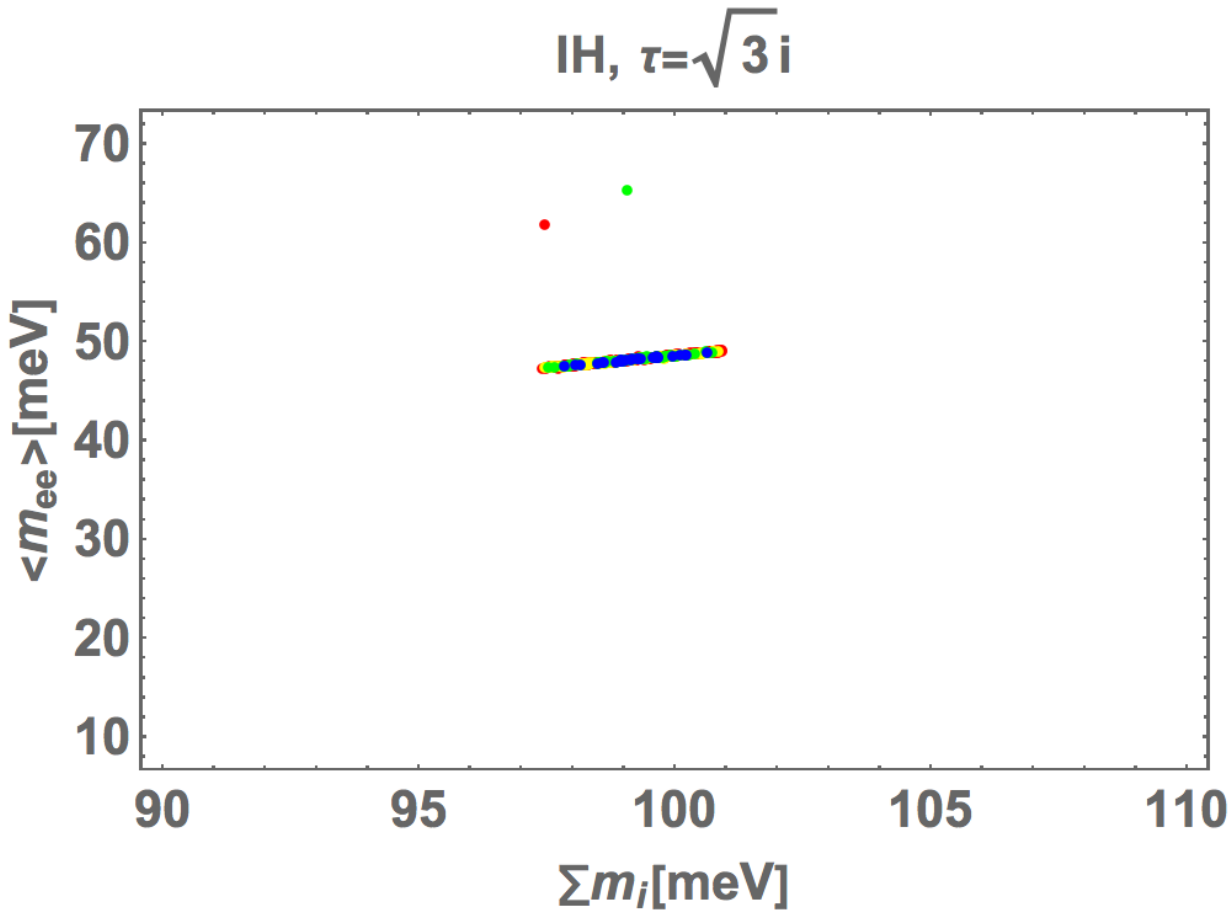}
   \includegraphics[scale=0.35]{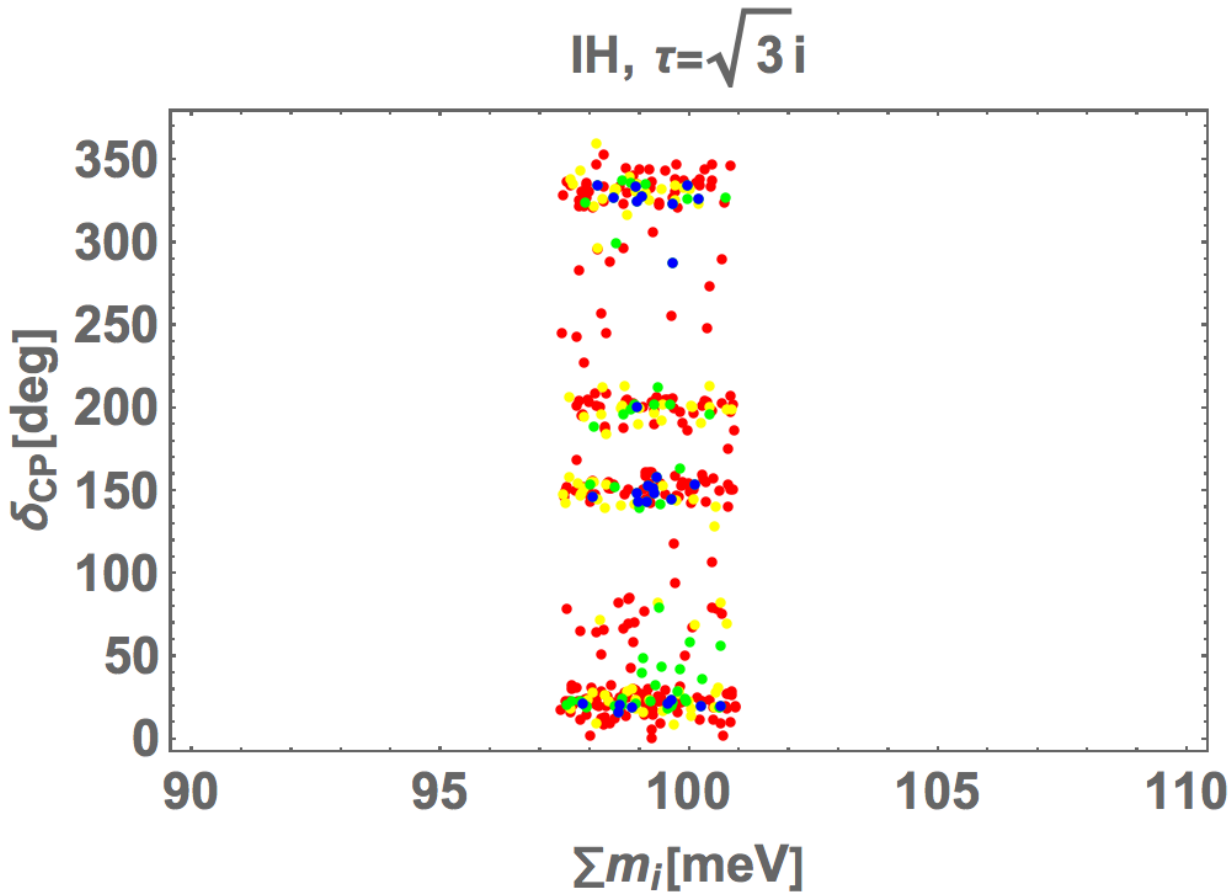}
 \caption{IH: Allowed regions in case of $\tau=\sqrt{3}i$.}
 \label{fig:r3i-ih}
\end{figure}
In Fig.~\ref{fig:r3i-ih}, we show the allowed region in case of $\tau=\sqrt{3}i$, where whole the legends are the same as the case of $\tau=i$. 
The left-top one implies that we find solutions at the exact special point $\tau=\sqrt{3}i$. 
The right-top one suggests that there exist several islands where any value is taken for $\delta_{\rm CP}$ even though there are several dense parts.
But, we have $0^\circ\lesssim \alpha_{21}\lesssim100^\circ$ and $240^\circ\lesssim \alpha_{21}\lesssim360^\circ$
The bottom ones tell us $\sum m_i=$97-101 meV that directly comes from the experimental value, while  $\langle m_{ee}\rangle$ tends to be localized at 46-50 meV.


\subsubsection{$\tau=-\frac14+\frac{\sqrt{15}}{4}i$}

\begin{figure}[H]
  \includegraphics[scale=0.35]{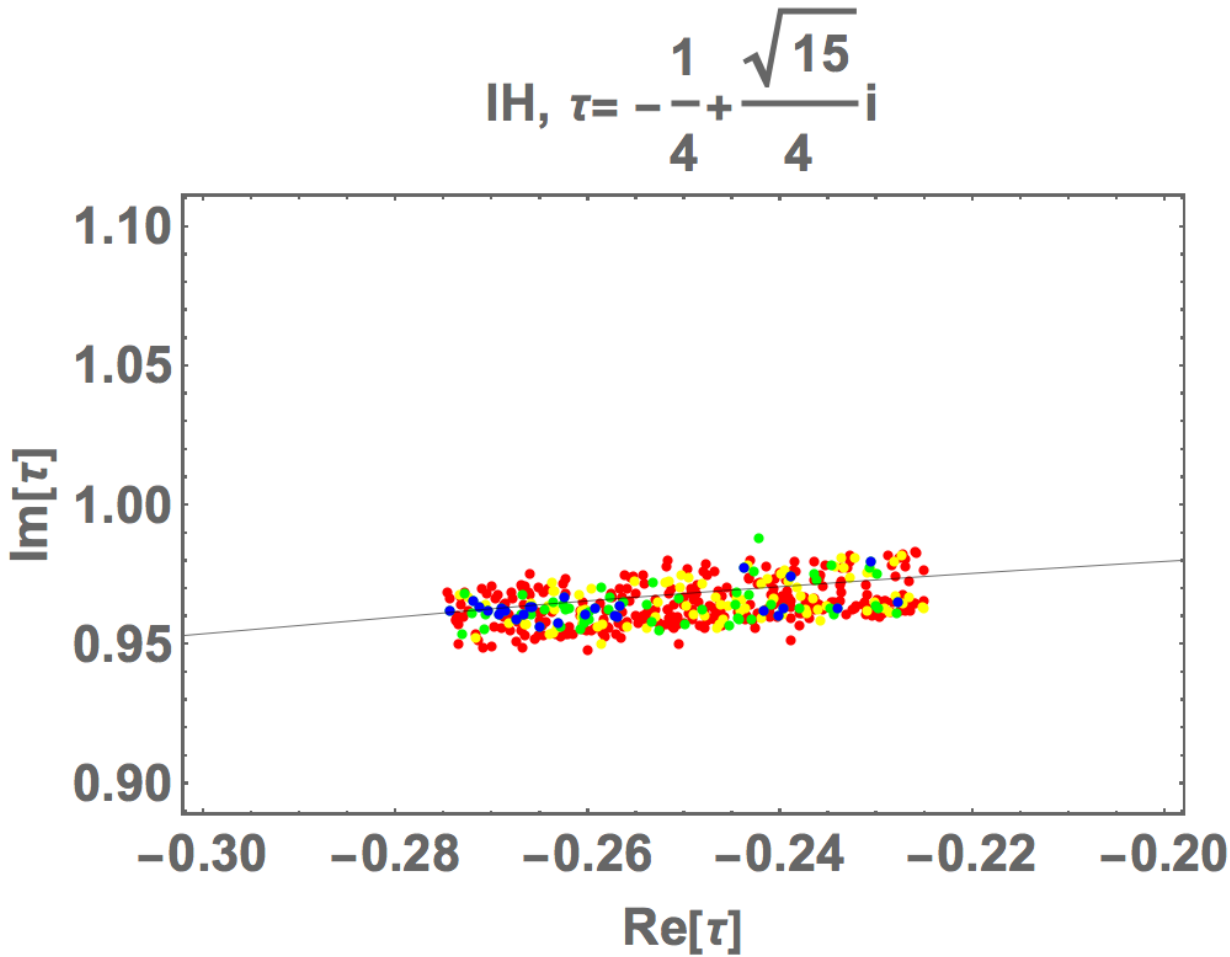}
  \includegraphics[scale=0.35]{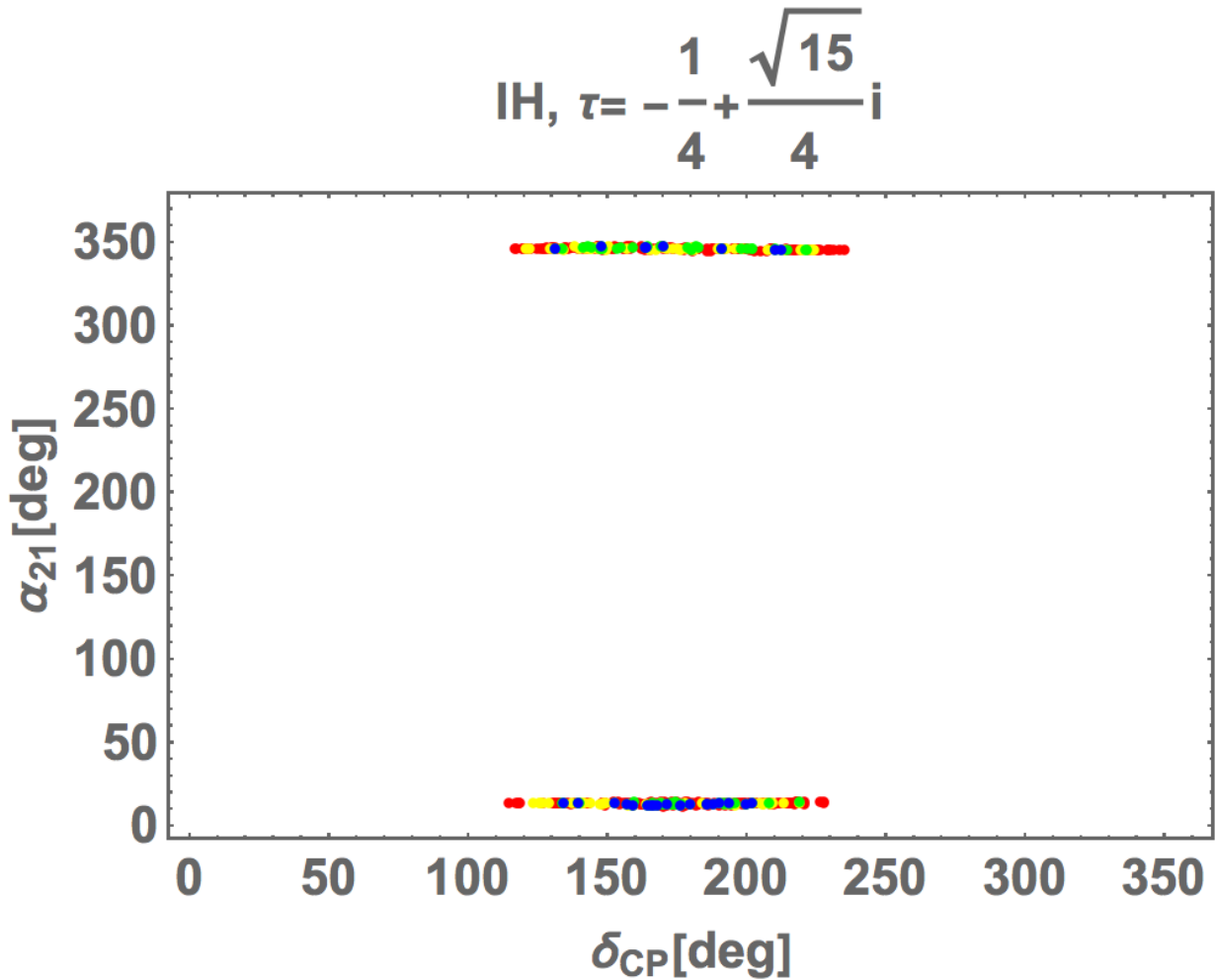}\\
   \includegraphics[scale=0.35]{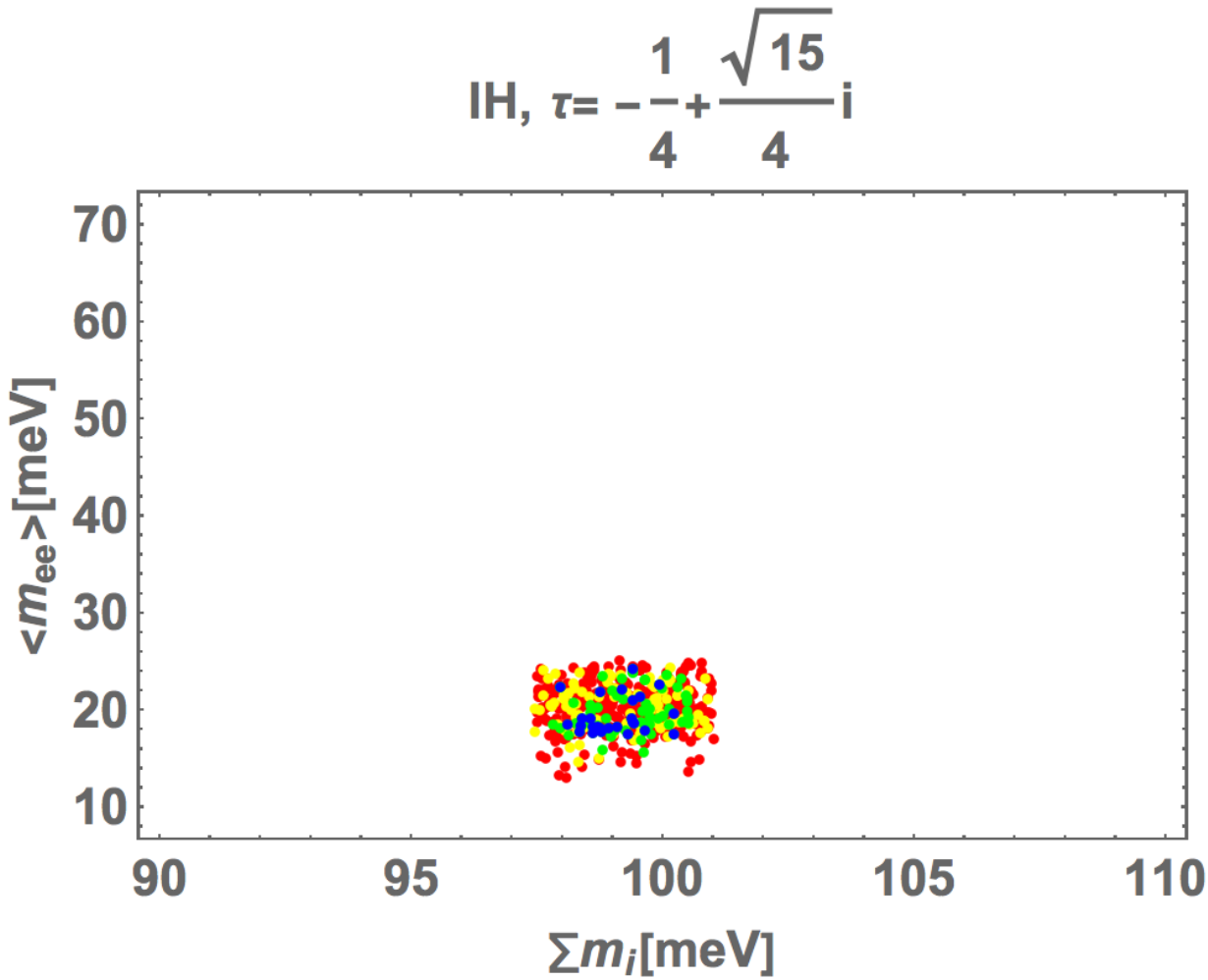}
   \includegraphics[scale=0.35]{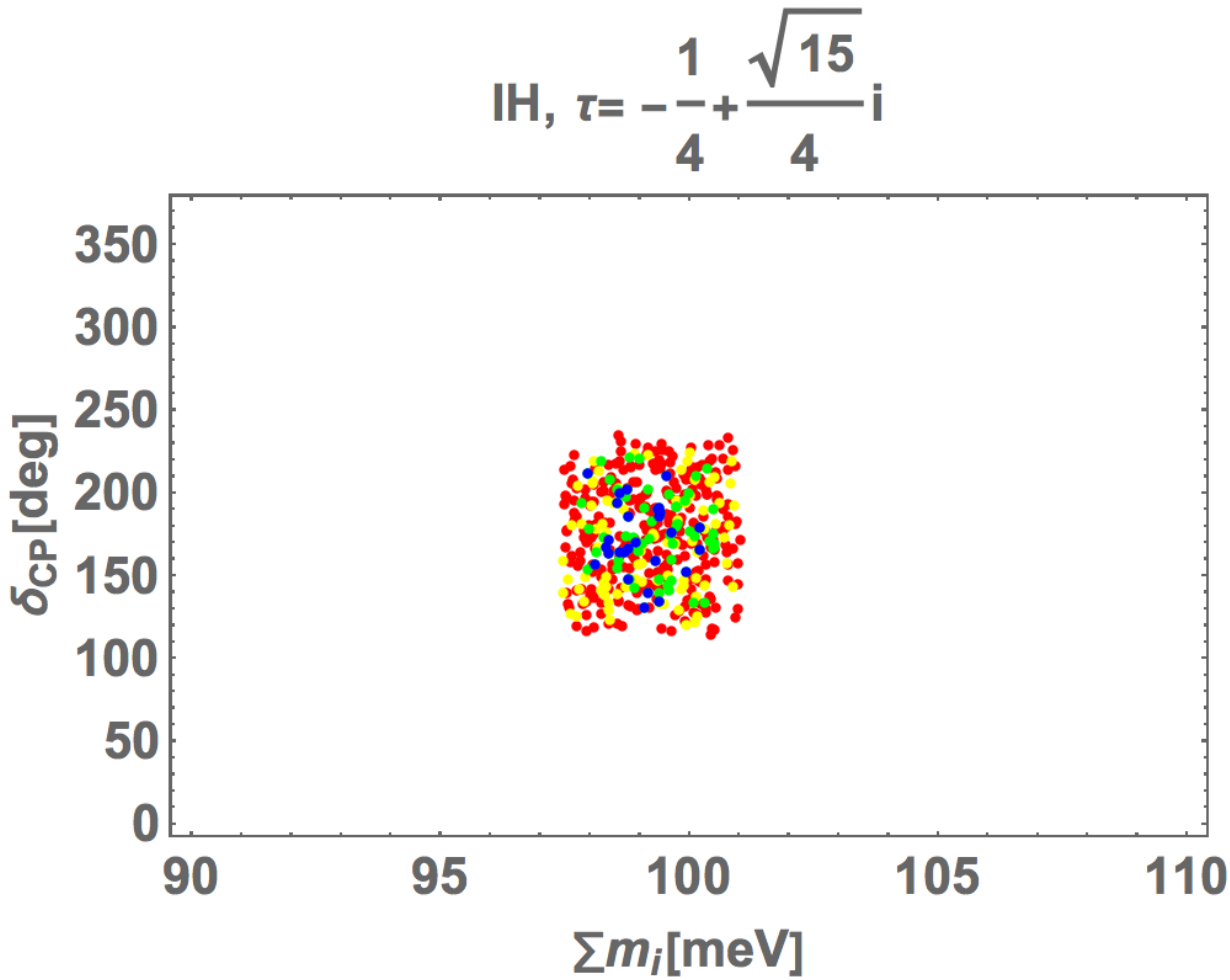}
 \caption{IH: Allowed regions in case of $\tau=-\frac14+\frac{\sqrt{15}}{4}i$.}
 \label{fig:14-ih}
\end{figure}
In Fig.~\ref{fig:14-ih}, we show the allowed region in case of $\tau=-\frac14+\frac{\sqrt{15}}{4}i$, where whole the legends are the same as the case of $\tau=i$. 
The left-top one implies that we find solutions at near by $\tau=-\frac14+\frac{\sqrt{15}}{4}i$. 
The right-top one suggests $\alpha_{21}\sim0^\circ$ and $110^\circ\lesssim \delta_{\rm CP}\lesssim 240^\circ$.
The bottom ones tell us 98 meV$\lesssim\sum m_i\lesssim$101 meV that directly comes from the experimental value, while  $\langle m_{ee}\rangle$ is localized at [12--24] meV.

 \subsection{Multiple dark gauge bosons in the model}

 Here we briefly discuss possible phenomenology regarding dark gauge bosons from $SU(2)_H$.
 In the model there are three massive new neutral gauge bosons from spontaneously broken hidden $SU(2)$ gauge symmetry.
 Mass terms of gauge fields are obtained by kinetic terms of scalar fields 
 \begin{align}
  L_{K} = & (D_\mu H_u)^\dagger (D^\mu H_u) + (D_\mu H_d)^\dagger (D^\mu H_d) + Tr[(D_\mu \Delta)^\dagger (D^\mu \Delta)]  \nonumber \\
& + (D_\mu H_X)^\dagger (D^\mu H_X)  + \sum_{i=1,2} Tr[(D_\mu \Phi_i)^\dagger (D^\mu \Phi_i)] \, .
\end{align}
Covariant derivatives are written by
\begin{align}
 D_\mu H_u = & \partial_\mu H_u - i g_2 W_\mu^i \frac{\sigma^i}{2} H_u - i g_1 \frac{1}{2} B_\mu H_u, \\
 D_\mu H_d = & \partial_\mu H_d - i g_2 W_\mu^i \frac{\sigma^i}{2} H_d + i g_1 \frac{1}{2} B_\mu H_d, \\
 D_\mu H_X = & \partial_\mu H_X - i g_X \frac{\sigma^i}{2} X_{\mu}^i H_X, \\
 D_\mu \Delta = & \partial_\mu \Delta - i g_H [\frac{\sigma^i}{2} W_{H_\mu}^i, \Delta], \\
 D_\mu \Phi_1 = & \partial_\mu \Phi_1 - i g_2 W_\mu^i \frac{\sigma^i}{2} \Phi_1 + i g_X \frac{\sigma^i}{2} \Phi_1 X_{\mu}^i - i g_1 \frac{1}{2} B_\mu \Phi_1, \\
  D_\mu \Phi_2 = & \partial_\mu \Phi_2 - i g_2 W_\mu^i \frac{\sigma^i}{2} \Phi_2 + i g_X \frac{\sigma^i}{2} \Phi_1 X_{\mu}^i + i g_1 \frac{1}{2} B_\mu \Phi_2, 
 \end{align}
 where $X^i_\mu (W^i_\mu)$ is the gauge associated with $SU(2)_{H(L)}$, $B_\mu$ is $U(1)_Y$ gauge field, and $\{g_1, g_2, g_X \}$ are respectively gauge couplings for $\{ U(1)_Y, SU(2)_L, SU(2)_H\}$. 
 After spontaneous gauge symmetry breaking, we obtain mass terms for gauge fields such that
 \begin{align}
 \mathcal{L}_M^{\rm gauge} = & \frac12 M_1^2 X^1_\mu X^{1 \mu} + \frac12 M_2^2 X^2_\mu X^{2 \mu} + \frac12 M_3^2 X^3_\mu X^{3 \mu} + \frac12 M_{\hat{Z}}^2 \hat{Z}_\mu \hat{Z}^{\mu} + M_W^2 W^+_\mu W^{- \mu} \nonumber \\
 & + M^2_{\hat{Z} X^1} X^1_\mu \hat{Z}^\mu + M^2_{\hat{Z} X^3} X^3_\mu \hat{Z}^\mu.
 \end{align}
 The coefficients of each term are written by
 \begin{align}
& M_1^2 = \frac{g_X}{4} (2v_X^2 + \bar \kappa^2 + \bar \zeta^2 + 4 (v_+ - v_-)^2), \nonumber \\
& M_2^2 = \frac{g_X}{4} (2v_X^2 + \bar \kappa^2 + \bar \zeta^2 + 4 (v_+ + v_-)^2), \nonumber \\
& M_3^2 = \frac{g_X}{4} (2v_X^2 + \bar \kappa^2 + \bar \zeta^2 + 4 (v_+^2 + v_-^2)), \nonumber \\
& M_W^2 = \frac{g^2_2}{4} (v^2 + \bar \kappa^2 + \bar \zeta^2),  \quad M_{\hat Z}^2 = \frac{g^2_Z}{4} (v^2 + \bar \kappa^2 + \bar \zeta^2), \nonumber \\ 
& M^2_{\hat{Z} X^1} = \frac{g_X g_Z}{2} (\kappa \kappa' - \zeta \zeta'), \quad M^2_{\hat{Z} X^3} = - \frac{g_X g_Z}{4} (\kappa^2 - \kappa'^2 + \zeta^2 - \zeta'^2),
 \end{align}
 where $v^2 = v_u^2 + v_d^2$, $\bar \kappa^2 = \kappa^2 + \kappa'^2$, $\bar \zeta^2 = \zeta^2 + \zeta'^2$, $\hat Z_\mu = \cos \theta_W W^3_\mu - \sin \theta_W B_\mu$ and $g_Z = \sqrt{g_1^2 + g_2^2}$.
 The masses of dark gauge bosons are $M_1 \sim M_2 \sim M_3 \sim g_X v_X/2$ for $v_X \gg \{ \kappa, \kappa', \zeta, \zeta', v_+, v_-  \}$.
 We also find that the SM $Z$ boson mixes with $X^1$ and $X^3$ while $X^2$ does not mix.
 The mass eigenstates and eigenvalues of neutral gauge bosons can be obtained by diagonalizing $3 \times 3$ mass matrix for $\{ \hat Z, X^1, X^3 \}$.
 Here we omit to explicitly diagonalize the mass matrix and just require the mixings are small so that the model is safe from electroweak precision tests.
 For discussion we write mass eigenstate as $\{Z, Z'_1, Z'_2, Z'_3 \}$ where $Z'_2 = X^2$ and $Z$ is the SM $Z$ boson.
 
 By small mixing effect, two dark gauge bosons $\{Z'_1, Z'_2 \}$ can decay into the SM particles directly.
 Their behavior is similar to dark photons that are tested by various experiments including current/future ones~\cite{Bauer:2018onh}. 
 On the other hand, $Z'_2$ can decay into the SM particles through off-shell $Z'_{1,3}$, and it would be a long-lived particle.
 Then $Z'_2$ can provide an interesting signature of the model at collider experiments.
  The $Z'_2$ can be produced through scalar mixing; for example $Z'_2$ can be obtained from the SM Higgs decay $h \to Z'_2 Z'_2$ if $m_h > 2 m_{Z'_2}$ via scalar mixing among
  $H_u$, $H_d$ and $H_X$. 
  Produced $Z'_2$ would be long lived and decay mode is such as $Z'_2 \to Z'^\ast_1 Z'^\ast_3 \to f_{SM} \bar{f}_{SM} f_{SM} \bar{f}_{SM}$ where $f_{SM}$ is a SM fermion.
 We thus expect long-lived particle decaying into 4 fermions, and the signal can be tested by collider experiments targeting to detect long-lived particles~\cite{Feng:2017uoz,Chou:2016lxi,Evans:2017lvd,Gligorov:2017nwh,Feng:2022inv}.
 Detailed discussion of collider signals is beyond the scope of this paper and we leave it as future work.

 \section{Summary and discussions}
We have studied flavor phenomenologies in a basis of a double covering of modular $A_4$
with a hidden $SU(2)$ symmetry, in which
we have worked on regions at novel moduli points in addition to well-studied fixed points in $SL(2,\mathbb{Z})$ moduli space. 
Since these special points of $SL(2,\mathbb{Z})$ moduli space were known to be statistically favored in flux compactifications of Type IIB string theory \cite{Ishiguro:2020tmo}, it is interesting to study the phenomenological implications of these unexplored moduli values for the lepton sector. 

In our model, the neutrino masses are approximately obtained by using the seesaw mechanism due to our two additional symmetries.
We have performed chi square numerical analysis for each of the fixed and special points in NH and IH and demonstrated predictions for each case. 
In the result we have obtained good predictions regarding $\sum m_i$ and $\langle m_{ee} \rangle$ due to model structure with $T'$ symmetry and constrained modulus $\tau$ at fixed/special points. 
In case of NH, we have found that there is no deviation from all the three special points which satisfy the neutrino oscillation, while two fixed points require deviations from the exact points to find experimental solutions. Thus, there is a tendency that allowed regions of special points are rather localized than the ones of fixed points at least in focusing on the chi square within 1$\sigma$ interval.\footnote{See, e.g., Ref. \cite{Ishiguro:2022pde}, for the mechanism to obtaining deviations from the fixed points.}
In case of IH, it provides $\langle m_{ee} \rangle$ prediction around testable region and some predictions of $\delta_{\rm CP}$ depending on different fixed/special points. While these predictions are still safe from the current constraints, they can be tested in future experiments; for example, future KamLAND-ZEN, nEXO and LEGEND for $\langle m_{ee} \rangle$, and CMB-S4~\cite{CMB-S4:2016ple} for $\sum m_i$. 
We thus have good testability in the neutrino sector, and future measurements may find favored fixed/special point. 
Furthermore, we have discussed detectability at collider physics via kinetic mixings from the hidden $SU(2)$ symmetry.

\acknowledgments

The work was supported by the Fundamental Research Funds for the Central Universities (T.~N.), JSPS KAKENHI Grant Numbers JP20K14477 (H. Otsuka), JP22J12877 (K. I.) and JP23H04512 (H. Otsuka). 

\section*{Appendix}

Here we summarize $T'$ modular forms that are used in the model. 
The lowest weight $T'$ modular form is weight 1 doublet written by
\begin{align}
Y_2^{(1)}(\tau) & = \begin{pmatrix} y_1(\tau) && y_2(\tau) \end{pmatrix}^T, \nonumber \\
y_1(\tau) & \simeq \sqrt{2} e^{\frac{7 \pi}{12} i} e^{2 \pi i \tau} (1+ e^{2\pi i \tau} + 2(e^{2\pi i \tau})^2 +2 (e^{2\pi i \tau})^4 + (e^{2\pi i \tau})^5 + 2 (e^{2\pi i \tau})^6), \nonumber \\
y_2(\tau) & \simeq \frac13 + 2 e^{2\pi i \tau} + 2(e^{2\pi i \tau})^3 +2 (e^{2\pi i \tau})^4 + 4(e^{2\pi i \tau})^7 + 2 (e^{2\pi i \tau})^9,
\end{align}
where we applied an approximated form for the $y_1$ and $y_2$. 
Higher weight modular forms are constructed by products of the lowest one. 

Modular weight 2 triplet one is given by
\begin{align}
& Y_{3}^{(2)}  = \begin{pmatrix} y^{(2)}_{1} && y^{(2)}_{2} &&  y^{(2)}_{3} \end{pmatrix}^T, \nonumber \\
& y^{(2)}_{1} = e^{\frac{\pi}{6} i} y_2^2, \quad 
y^{(2)}_{2} = e^{\frac{7 \pi}{12} i} y_1 y_2, \quad
y^{(2)}_{3} = y_1^2,
\end{align}
where $\tau$ dependence of modular form is omitted; also in the equations below.
Modular weight 3 doublets are given by
\begin{align}
& Y_{2_1}^{(3)}  = \begin{pmatrix} y^{(3)}_{1} && y^{(3)}_{2} && \end{pmatrix}^T, \quad 
Y^{(3)}_{2_2}  = \begin{pmatrix} y'^{(3)}_{1} && y'^{(3)}_{2} && \end{pmatrix}^T,\nonumber\\
& y^{(3)}_{1} = 3 e^{\frac{\pi}{6} i} y_1 y_2^2, \quad 
y^{(3)}_{2} = \sqrt{2} e^{\frac{5 \pi}{12} i} y_1^3 - e^{\frac{\pi}{6} i} y_2^3, \nonumber \\ 
& y'^{(3)}_{1} = y_1^3 +(1-i)y_2^3, \quad 
y'^{(3)}_{2} = -3 y_2 y_1^2.
\end{align}
Modular weight 4 triplet one is given by
\begin{align}
& Y_{3}^{(4)}  = \begin{pmatrix} y^{(4)}_{1} && y^{(4)}_{2} &&  y^{(4)}_{3} \end{pmatrix}^T, \nonumber \\
& y^{(4)}_{1} = \sqrt{2} e^{\frac{7 \pi}{12} i} y_1^3 y_2 - e^{\frac{\pi}{3} i} y_2^4, \quad 
y^{(4)}_{2} =-y_1^4 -(1-i) y_1 y_2^3, \quad
y^{(4)}_{3} = 3 e^{\frac{\pi}{6} i} y_1^2 y_2^2.
\end{align}
%
%
Modular weight 6 triplets are given by
\begin{align}
& Y_{3_1}^{(6)}  = \begin{pmatrix} y^{(6)}_{1} && y^{(6)}_{2} &&  y^{(6)}_{3} \end{pmatrix}^T, \quad
 Y_{3_2}^{(6)}  = \begin{pmatrix} y'^{(6)}_{1} && y'^{(6)}_{2} &&  y'^{(6)}_{3} \end{pmatrix}^T, \nonumber \\
& y^{(6)}_1 = (4 y_1^3 y_2 + (1-i)y_2^4) y^{(2)}_1, \quad 
y^{(6)}_2 = (4 y_1^3 y_2 + (1-i)y_2^4) y^{(2)}_2, \quad
y^{(6)}_3 = (4 y_1^3 y_2 + (1-i)y_2^4) y^{(2)}_3, \nonumber \\
& y'^{(6)}_1 = ((1+i) y_1^4 - 4 y_1 y_2^3) y^{(2)}_3, \quad
y'^{(6)}_2 = ((1+i) y_1^4 - 4 y_1 y_2^3) y^{(2)}_1, \quad
y'^{(6)}_3 = ((1+i) y_1^4 - 4 y_1 y_2^3) y^{(2)}_2.
\end{align}
Modular weight 7 doublets are given by
\begin{align}
& Y_{2_1}^{(7)}  = \begin{pmatrix} f^{(7)}_{1} && f^{(7)}_{2} && \end{pmatrix}^T, \
Y_{2_2}^{(7)}  = \begin{pmatrix} g^{(7)}_{1} && g^{(7)}_{2} && \end{pmatrix}^T, \ 
Y_{2'}^{(7)}  = \begin{pmatrix} f'^{(7)}_{1} && f'^{(7)}_{2} && \end{pmatrix}^T, \ 
Y_{2''}^{(7)}  = \begin{pmatrix} f''^{(7)}_{1} && f''^{(7)}_{2} && \end{pmatrix}^T, \nonumber \\
& f^{(7)}_1 = (4 y_1^3 y_1 + (1-i) y_2^4) y^{(3)}_1, \quad
f^{(7)}_2 = (4 y_1^3 y_1 + (1-i) y_2^4) y^{(3)}_2, \nonumber \\
& g^{(7)}_1 = ((1+i) y_1^4  -4 y_1 y_2^3) y'^{(3)}_1, \quad
g^{(7)}_2 = ((1+i) y_1^4  -4 y_1 y_2^3) y'^{(3)}_2, \nonumber \\
& f'^{(7)}_1 = ((1+i) y_1^4  -4 y_1 y_2^3) y^{(3)}_1, \quad
f'^{(7)}_2 = ((1+i) y_1^4  -4 y_1 y_2^3) y^{(3)}_2, \nonumber \\
& f''^{(7)}_1 = (4 y_1^3 y_1 + (1-i) y_2^4) y'^{(3)}_1, \quad
f''^{(7)}_2 = (4 y_1^3 y_1 + (1-i) y_2^4) y'^{(3)}_2.
\end{align}
Modular weight 10 triplets are given by
\begin{align}
& Y_{3_1}^{(10)}  = \begin{pmatrix} y^{(10)}_{1} && y^{(10)}_{2} &&  y^{(10)}_{3} \end{pmatrix}^T, \
Y_{3_2}^{(10)}  = \begin{pmatrix} y'^{(10)}_{1} && y'^{(10)}_{2} &&  y'^{(10)}_{3} \end{pmatrix}^T, \
Y_{3_3}^{(10)}  = \begin{pmatrix} y''^{(10)}_{1} && y''^{(10)}_{2} &&  y''^{(10)}_{3} \end{pmatrix}^T, \nonumber \\
& y^{(10)}_1 = (4 y_1^3 y_2+ (1-i)y_2^4)^2 y^{(2)}_1, \quad
y^{(10)}_2 = (4 y_1^3 y_2+ (1-i)y_2^4)^2 y^{(2)}_2, \quad
y^{(10)}_3 = (4 y_1^3 y_2+ (1-i)y_2^4)^2 y^{(2)}_3, \nonumber \\
& y'^{(10)}_1 = (4 y_1^3 y_2+ (1-i)y_2^4)((1+i)y_1^4 -4y_1y_2^3) y^{(2)}_3, \quad
y'^{(10)}_2 = (4 y_1^3 y_2+ (1-i)y_2^4)((1+i)y_1^4 -4y_1y_2^3) y^{(2)}_1, \nonumber \\
& y'^{(10)}_1 = (4 y_1^3 y_2+ (1-i)y_2^4)((1+i)y_1^4 -4y_1y_2^3)  y^{(2)}_2, \nonumber \\
& y''^{(10)}_1 = ((1+i)y_1^4 -4y_1y_2^3)^2 y^{(2)}_2, \quad
y''^{(10)}_2 = ((1+i)y_1^4 -4y_1y_2^3)^2 y^{(2)}_3, \quad 
y''^{(10)}_3 = ((1+i)y_1^4 -4y_1y_2^3)^2 y^{(2)}_1.
\end{align}


\bibliography{references}{}
\bibliographystyle{JHEP} 

\end{document}